\documentclass{emulateapj}
\usepackage{natbib,amsmath,enumitem}
\usepackage{epsfig}
\usepackage{color}

\newcommand{\acounits}{\mbox{M$_\odot$ pc$^{-2}$ (K km s$^{-1}$)$^{-1}$}}

\newcommand{\sdavg}{\mbox{$\left< \Sigma_{\rm 40pc} \right>$}}
\newcommand{\bavg}{\mbox{$\left< b_{\rm 40pc} \right>$}}
\newcommand{\rhoavg}{\mbox{$\left< \rho_{\rm 40pc} \right>$}}
\newcommand{\tffavg}{\mbox{$\left< \tau_{\rm ff,40pc} \right>$}}
\newcommand{\sigavg}{\mbox{$\left< \sigma_{\rm 40pc} \right>$}}
\newcommand{\effavg}{\mbox{$\left< \epsilon_{\rm ff,40pc} \right>$}}
\newcommand{\tffsdavg}{\mbox{$\tau_{\rm ff} \left(\sdavg \right)$}}
\newcommand{\effsdavg}{\mbox{$\epsilon_{\rm ff} \left(\sdavg \right)$}}
\newcommand{\rhosdavg}{\mbox{$\rho \left(\sdavg \right)$}}

\shorttitle{Cloud Scale ISM Structure and Star Formation in M51}
\shortauthors{Leroy et al.}

\begin{document}

\slugcomment{Accepted for Publication in the Astrophysical Journal} 
\title{Cloud Scale ISM Structure and Star Formation in M51}

\author{
Adam~K.~Leroy\altaffilmark{1},
Eva~Schinnerer\altaffilmark{2},
Annie~Hughes\altaffilmark{3,4},
J.~M.~Diederik~Kruijssen\altaffilmark{5,2},
Sharon~Meidt\altaffilmark{2},
Andreas~Schruba\altaffilmark{6},
Jiayi~Sun\altaffilmark{1},
Frank~Bigiel\altaffilmark{7},
Gonzalo~Aniano\altaffilmark{8},
Guillermo~A.~Blanc\altaffilmark{9,10,11},
Alberto~Bolatto\altaffilmark{12},
M\'{e}lanie~Chevance\altaffilmark{5},
Dario~Colombo\altaffilmark{13},
Molly~Gallagher\altaffilmark{1},
Santiago~Garcia-Burillo\altaffilmark{14},
Carsten~Kramer\altaffilmark{15},
Miguel~Querejeta\altaffilmark{16},
Jerome~Pety\altaffilmark{17,18},
Todd~A.~Thompson\altaffilmark{1,19},
Antonio~Usero\altaffilmark{14}
}
\altaffiltext{1}{Department of Astronomy, The Ohio State University, 140 West 18th Avenue, Columbus, OH 43210}
\altaffiltext{2}{Max-Planck-Institut f\"ur Astronomie, K\"onigstuhl 17, 69117, Heidelberg, Germany}
\altaffiltext{3}{CNRS, IRAP, 9 av. du Colonel Roche, BP 44346, F-31028 Toulouse cedex 4, France}
\altaffiltext{4}{Universit\'{e} de Toulouse, UPS-OMP, IRAP, F-31028 Toulouse cedex 4, France}
\altaffiltext{5}{Astronomisches Rechen-Institut, Zentrum f\"{u}r Astronomie der Universit\"{a}t Heidelberg, M\"{o}nchhofstra\ss e 12-14, 69120 Heidelberg, Germany}
\altaffiltext{6}{Max-Planck-Institut f\"ur extraterrestrische Physik, Giessenbachstra{\ss}e 1, 85748 Garching, Germany}
\altaffiltext{7}{Institute f\"ur theoretische Astrophysik, Zentrum f\"ur Astronomie der Universit\"at Heidelberg, Albert-Ueberle Str. 2, 69120 Heidelberg, Germany.}
\altaffiltext{8}{Princeton University Observatory, Peyton Hall, Princeton, NJ 08544-1001}
\altaffiltext{9}{Departamento de Astronom\'{\i}a, Universidad de Chile, Casilla 36-D, Santiago, Chile}
\altaffiltext{10}{Centro de Astrof\'{\i}sica y Tecnolog\'{\i}as Afines (CATA), Camino del Observatorio 1515, Las Condes, Santiago, Chile}
\altaffiltext{11}{Visiting Astronomer, Observatories of the Carnegie Institution for Science, 813 Santa Barbara St, Pasadena, CA, 91101, USA}
\altaffiltext{12}{Department of Astronomy, Laboratory for Millimeter-wave Astronomy, and Joint Space Institute, University of Maryland, College Park, Maryland 20742, USA}
\altaffiltext{13}{Max-Planck-Institut f\"ur Radioastronomie, Auf dem H\"ugel 69, D-53121 Bonn, Germany}
\altaffiltext{14}{Observatorio Astron\'omico Nacional (IGN), C/ Alfonso XII, 3, 28014 Madrid, Spain}
\altaffiltext{15}{Instituto Radioastronom\'{\i}a Milim\'{e}trica (IRAM), Av. Divina Pastora 7, Nucleo Central, E-18012 Granada, Spain}
\altaffiltext{16}{European Southern Observatory, Karl-Schwarzschild-Stra{\ss}e 2, D-85748 Garching, Germany}
\altaffiltext{17}{Institut de Radioastronomie Millim\`{e}trique (IRAM), 300 Rue de la Piscine, F-38406 Saint Martin d'H\`{e}res, France}
\altaffiltext{18}{Observatoire de Paris, 61 Avenue de l'Observatoire, F-75014 Paris, France}
\altaffiltext{19}{Center for Cosmology \& Astro-Particle Physics, The Ohio State University, Columbus, OH 43210}

\begin{abstract}
We compare the structure of molecular gas at $40$~pc resolution to the ability of gas to form stars across the disk of the spiral galaxy M51. We break the PAWS survey into $370$~pc and $1.1$~kpc resolution elements, and within each we estimate the molecular gas depletion time ($\tau_{\rm Dep}^{\rm mol}$), the star formation efficiency per free fall time ($\epsilon_{\rm ff}$), and the mass-weighted cloud-scale (40~pc) properties of the molecular gas: surface density, $\Sigma$, line width, $\sigma$, and $b\equiv\Sigma/\sigma^2\propto\alpha_{\rm vir}^{-1}$, a parameter that traces the boundedness of the gas. We show that the cloud-scale surface density appears to be a reasonable proxy for mean volume density. Applying this, we find a typical star formation efficiency per free-fall time, $\effsdavg \sim 0.3{-}0.36\%$, lower than adopted in many models and found for local clouds. More, the efficiency per free fall time anti-correlates with both $\Sigma$ and $\sigma$, in some tension with turbulent star formation models.  The best predictor of the rate of star formation per unit gas mass in our analysis is $b \equiv \Sigma / \sigma^2$, tracing the strength of self gravity, with $\tau_{\rm Dep}^{\rm mol} \propto b^{-0.9}$. The sense of the correlation is that gas with stronger self-gravity (higher $b$) forms stars at a higher rate (low $\tau_{\rm Dep}^{\rm mol}$). The different regions of the galaxy mostly overlap in $\tau_{\rm Dep}^{\rm mol}$ as a function of $b$, so that low $b$ explains the surprisingly high $\tau_{\rm Dep}^{\rm mol}$ found towards the inner spiral arms found by by Meidt et al. (2013). 
\end{abstract}


\section{Introduction}
\label{sec:intro}

In the local universe, star formation occurs in molecular gas. The recent star formation rate (SFR) correlates better with tracers of molecular gas than tracers of atomic gas \citep{SCHRUBA11,BLANC09,BIGIEL08,LEROY08}, even though atomic gas represents the dominant reservoir by mass of the interstellar medium (ISM) in galaxies at $z=0$ \citep[e.g.,][]{SAINTONGE11}. But even within the molecular ISM of a galaxy, only a small fraction of the gas participates in star formation at any given time \citep[e.g.,][]{HEIDERMAN10,LADA10}, and the properties of molecular gas vary among galaxies and among regions within galaxies \citep[e.g.,][]{HUGHES13B,LEROY16}. The SFR per unit molecular gas mass should depend on these properties: e.g., the density, turbulence, and balance of potential and kinetic energy. As a result, we  expect star formation to proceed at different specific (per unit gas mass) rates in different environments.

Observations indeed indicate that the SFR per unit molecular gas mass does vary across the local galaxy population \citep[][]{YOUNG96}. High stellar mass, early type galaxies show comparatively low SFRs per unit H$_2$ mass \citep{SAINTONGE11,LEROY13,DAVIS14}. Starbursts, especially galaxy-wide bursts induced by major galaxy mergers, have a high SFR per unit H$_2$ mass \citep[e.g.,][]{KENNICUTT98B,GAO04}. So do some galaxy centers \citep[e.g.,][]{LEROY13,LEROY15A}. Low stellar mass, low metallicity, late-type galaxies exhibit a high SFR per unit CO emission \citep[e.g.,][]{YOUNG96,LEROY13,SCHRUBA12,SCHRUBA17}. Although the translation of CO emission into H$_2$ mass remains uncertain in these systems \citep[][]{BOLATTO13B}, several works argue that the SFR per H$_2$ mass is indeed higher in these systems \citep[e.g.,][]{GARDAN07,BOTHWELL14,HUNT15}. Within galaxies, dynamical effects can both enhance \citep{KODA09,SUWANNAJAK14} and suppress \citep{MEIDT13} the SFR-per-H$_2$. As our ability to observe the molecular ISM across diverse environments improves, the list of observed variations in the SFR per unit H$_2$ mass continues to grow.

Though driven by large-scale environmental factors, the observed SFR-per-H$_2$ variations must have their immediate origins in the properties of the clouds that host star formation. That is, in an environment with a high SFR per unit gas mass, we expect the configuration and small-scale physical properties of the molecular ISM to be more conducive to star formation.

Recent theoretical work exploring variations in SFR-per-H$_2$ has focused on the properties of turbulent molecular clouds. In such models, the mean density, gravitational boundedness, and Mach number of a cloud determine its normalized rate of star formation \citep[e.g.,][]{PADOAN02,KRUMHOLZ05,PADOAN11,HENNEBELLE11,FEDERRATH12,FEDERRATH13}. These properties set the density structure of the cloud and the balance between kinetic and potential energy, determining the fraction of the gas in a directly star-forming, self-gravitating component. In such models, the gravitational free-fall time, $\tau_{\rm ff} \propto \rho^{-0.5}$, often emerges as the characteristic timescale for star formation at many scales \citep[e.g.,][]{KRUMHOLZ05}, albeit with a low efficiency per $\tau_{\rm ff}$ \citep[see also][]{MCKEE07}.

Observations and theory suggest that the turbulent motions in molecular clouds are driven at about at the scale of an individual cloud \citep[$d \approx 30{-}100$~pc, e.g., ][]{BRUNT03,MACLOW04,BRUNT09} making this the relevant scale for many of the models referenced above. Current millimeter-wave telescopes can observe the structure of molecular gas at these scales across large areas of galaxies. This allows the prospect to measure how the cloud-scale structure of the cold ISM relates to the ability of gas to form stars in different galactic environments.

In this paper, we carry out such a study targeting  M51. Our key data set is the PdBI Arcsecond Whirlpool Survey\footnote{This work is based on observations carried out with the IRAM NOEMA Interferometer and the IRAM 30-m telescope. IRAM is supported by INSU/CNRS (France), MPG (Germany) and IGN (Spain).} \citep[PAWS,][]{SCHINNERER13}. PAWS mapped CO~\mbox{(1-0)} emission from the inner $9 \times 6$~kpc of M51 at $40$~pc resolution \citep[adopting a distance of 7.6\,Mpc;][]{FELDMEIER97,CIARDULLO02}. From PAWS, we know the structure of the turbulent ISM at the scale of an individual giant molecular cloud \citep[GMC, see][]{HUGHES13B,HUGHES13A,COLOMBO14A}. Combining this information with infrared maps from {\em Herschel} and {\em Spitzer} \citep{MENTUCHCOOPER12,KENNICUTT03}, we measure how the cloud-scale structure of the ISM relates to M51's ability to form stars.

This analysis builds on studies by \cite{KODA09}, \citet{HUGHES13A,HUGHES13B}, and \citet{COLOMBO14A}, which showed that the cloud-scale ISM structure in M51 depends on environment. We also follow \citet{MEIDT13}, \citet{LIU11}, \citet{MOMOSE13}, and \citet{SHETTY13}, who compare gas and star formation in M51 and came to apparently contradictory conclusions regarding whether star formation proceeds more quickly or more slowly in the highest density regions. In particular, we follow \citet{MEIDT13} who also compared PAWS to infrared (IR) data, focusing on the impact of dynamics on the ability of gas to form stars.

We use the methodology described by \citet{LEROY16}. In this approach, we calculate the molecular gas depletion time, $\tau_{\rm Dep}^{\rm mol} \equiv M_{\rm mol}/{\rm SFR}$, averaged over a moderate-sized area, $\theta = 370{-}1100$~pc, and compare this to the mass-weighted $40$~pc surface density, line width, and self-gravity (virial parameter) with in the larger beam. This approach captures both ensemble averages and local physical conditions. We expect that $\tau_{\rm Dep}^{\rm mol}$ becomes well-defined only after averaging over an ensemble of star-forming regions in different evolutionary states \citep[e.g., see][]{SCHRUBA10,KRUIJSSEN14}. Meanwhile the beam-by-beam 40~pc structural measurements from PAWS allow us to test expectations from turbulent theories. By taking the mass-weighted average within each larger beam, we preserve the small scale structural information in the PAWS map.

\section{Methods}
\label{sec:methods}

\begin{deluxetable*}{cccccccccccc}
\tabletypesize{\scriptsize}
\tablecaption{Cloud Scale Structure, IR, and CO in M51 \label{tab:data}}
\tablewidth{0pt}
\tablehead{
\colhead{R.A.} & 
\colhead{Dec.} &
\colhead{Beam} &
\colhead{$r_{\rm gal}$\tablenotemark{a}} &
\colhead{$\Sigma_{\rm mol}$\tablenotemark{b}} &
\colhead{$\Sigma_{\rm SFR}$\tablenotemark{c}} &
\colhead{$\sdavg$\tablenotemark{b}} &
\colhead{$\sigavg$} &
\colhead{$\bavg$\tablenotemark{b}} & 
\colhead{$f_{\rm arm}$\tablenotemark{d}} & 
\colhead{$f_{\rm ia}$\tablenotemark{d}} & 
\colhead{$f_{\rm ctr}$\tablenotemark{d}}
\\
\colhead{($\arcdeg$)} & 
\colhead{($\arcdeg$)} &
\colhead{($\arcsec$)} &
\colhead{(kpc)} &
\colhead{(M$_\odot$~pc$^{-2}$)} &
\colhead{($\frac{{\rm M}_\odot~{\rm yr}^{-1}}{{\rm kpc}^{2}}$)} &
\colhead{(M$_\odot$~pc$^{-2}$)} &
\colhead{(km~s$^{-1}$)} &
\colhead{$\left(\frac{{\rm M}_\odot~{\rm pc}^{-2}}{({\rm km~s}^{-1})^{2}}\right)$} &
\colhead{(~)} &
\colhead{(~)} &
\colhead{(~)} 
\\
\colhead{(1)} & 
\colhead{(2)} &
\colhead{(3)} &
\colhead{(4)} &
\colhead{(5)} &
\colhead{(6)} &
\colhead{(7)} &
\colhead{(8)} &
\colhead{(9)} &
\colhead{(10)} &
\colhead{(11)} &
\colhead{(12)}
}
\startdata
 202.46964 &   47.19517 & 30 & 0.0 & 194.1 & 0.1928 & 380.8 & 10.4 & 3.53 & 0.01 & 0.00 & 0.98 \\ 
 202.46718 &   47.19806 & 30 & 0.4 & 204.3 & 0.1909 & 445.5 & 10.5 & 4.03 & 0.02 & 0.02 & 0.97 \\ 
 202.47209 &   47.19228 & 30 & 0.4 & 179.2 & 0.1688 & 353.2 &  9.9 & 3.57 & 0.03 & 0.01 & 0.96 \\ 
 202.46718 &   47.19228 & 30 & 0.5 & 200.6 & 0.1698 & 412.8 & 11.1 & 3.37 & 0.07 & 0.00 & 0.92 \\ 
 202.47209 &   47.19806 & 30 & 0.5 & 178.0 & 0.1749 & 368.2 &  9.8 & 3.81 & 0.05 & 0.01 & 0.94 \\ 
 202.46474 &   47.19517 & 30 & 0.5 & 223.7 & 0.1897 & 477.4 & 11.5 & 3.60 & 0.03 & 0.01 & 0.96 \\ 
 202.47453 &   47.19517 & 30 & 0.5 & 180.7 & 0.1775 & 365.5 & 10.2 & 3.52 & 0.02 & 0.00 & 0.97 \\ 
 202.46964 &   47.18940 & 30 & 0.8 & 170.8 & 0.1339 & 367.6 & 10.4 & 3.40 & 0.15 & 0.02 & 0.83 \\ 
 202.46964 &   47.20094 & 30 & 0.8 & 165.1 & 0.1493 & 394.9 &  9.6 & 4.33 & 0.10 & 0.04 & 0.86 \\ 
 202.47699 &   47.19228 & 30 & 0.8 & 161.7 & 0.1482 & 383.6 & 10.3 & 3.62 & 0.02 & 0.02 & 0.96 \\ 
\nodata & \nodata & \nodata & \nodata & \nodata & \nodata & \nodata & \nodata & \nodata & \nodata & \nodata & \nodata
\enddata
\tablenotetext{a}{Galactocentric radius for a thin disk and the orientation parameters quoted in Section~\ref{sec:convtophys}.}
\tablenotetext{b}{Molecular mass linearly translated from CO surface brightness using $\alpha_{\rm CO}=4.35$~\acounits .}
\tablenotetext{c}{Here, the SFR is a linear transformation of the TIR emission. See Section~\ref{sec:convtophys}.}
\tablenotetext{d}{Fraction of the CO flux in the beam that arises from arm, interarm, or central regions as defined following \citet{COLOMBO14A}.}
\tablecomments{The full version of this table is available as online material. The following uncertainties apply: (a) uncertainty in the distance, ${\sim}10\%$, linearly affects $r_{\rm gal}$, (b) for $\Sigma_{\rm mol}$ a ${\sim}10\%$ gain uncertainty applies to both resolutions, the statistical noise is on average $2.25$~M$_\odot$~pc$^{-2}$ at $10\arcsec$ resolution and $0.5$~M$_\odot$~pc$^{-2}$ at $30\arcsec$ resolution, (c) calibration uncertainties are of order $5{-}10$\%, multiband TIR estimates from \citet{GALAMETZ13} uncertain by ${\sim}0.08$~dex, translation from 70 $\mu$m to TIR scatters by an additional ${\sim}0.05$~dex, and statistical noise is ${\sim} 2.5 \times 10^6$~L$_\odot$~kpc$^{-2}$ ($\approx 4 \times 10^{-4}$~M$_\odot$~yr$^{-1}$~kpc$^{-2}$) at $30\arcsec$ and ${\sim} 9.4 \times 10^6$~L$_\odot$~kpc$^{-2}$ at $10\arcsec$ resolution ($\sim 1.3 \times 10^{-4}$ M$_\odot$~yr$^{-1}$~kpc$^{-2}$), (d) from our Monte Carlo calculation, typical statistical uncertainties in $\sdavg$, $\sigavg$, and $\bavg$ are $1.5\%$, $2\%$, and $3\%$ at $30\arcsec$ resolution and $4\%$, $5\%$, and $6\%$ at $10\arcsec$ resolution. Covariance in uncertainty at both resolutions is about $0.7$ between $\sdavg$ and $\sigavg$, $-0.4$ between $\sdavg$ and $\bavg$, and $-0.9$ between $\sigavg$ and $\bavg$. The ${\sim}10\%$ gain uncertainty also applies to $\sdavg$ and $\bavg$. These uncertainties do not account for translation to physical quantities. Selection criteria: at $30\arcsec$, we include all lines of sight where at least 50\% of the beam lies in the PAWS field. At $10\arcsec$, we include all lines of sight where 95\% of the beam lies in the PAWS field, $\Sigma_{\rm mol} > 5$~M$_\odot$~pc$^{-2}$, and $\Sigma_{\rm SFR} > 7.5 \times 10^{-3}$~M$_\odot$~yr$^{-1}$~kpc$^{-2}$.}
\end{deluxetable*}

\begin{figure}
\plotone{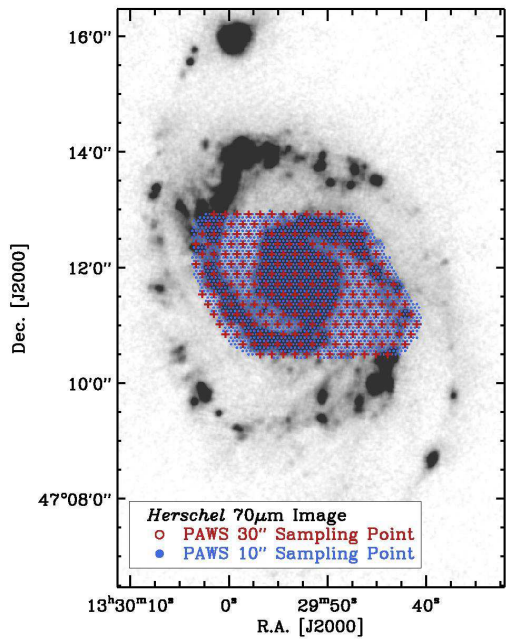}
\caption{The sampling points used in this paper, overlaid on the {\em Herschel} 70$\mu$m image of M51 \citep{MENTUCHCOOPER12}. The points are hexagonally packed and spaced by $15\arcsec$ (red) and $5\arcsec$ (blue), corresponding to half-beam spacing for our two working resolutions. The area studied is set by the PAWS field \citep{SCHINNERER13}. The coarser resolution $30\arcsec$ data allow the inclusion of all the IR bands.}
\label{fig:sampling}
\end{figure}

We wish to measure how small-scale ISM structure relates to the ability of gas to form stars in M51. To do this, we require region-by-region estimates of the recent star formation rate, the molecular gas reservoir, and the structure of molecular gas on the scale of an individual cloud. Using these, we correlate the cloud-scale structure of the molecular gas with the star formation rate per unit gas mass, expressed as a gas depletion time, $\tau_{\rm Dep}^{\rm mol}$.

We estimate these quantities and conduct a correlation analysis at $30\arcsec$ and $10\arcsec$ resolution. These correspond to linear resolutions of ${\sim} 1100$~pc and ${\sim} 370$~pc at our adopted distance of $7.6$~Mpc to M51 \citep{FELDMEIER97,CIARDULLO02}. At $30\arcsec$ resolution, we are able to include more IR bands in our SFR estimate. At $10\arcsec$ resolution, we are better able to resolve the dynamical features that drive the differences within the M51 cloud population \citep{KODA09,MEIDT13,MEIDT15}. At resolutions finer than $10\arcsec$, we cannot include infrared (IR) emission, our main SFR indicator (see appendix).

The choice of a few hundred pc to a kpc scale also ensures that within a resolution element we average over many individual star-forming regions. This allows us to avoid most effects related to the time evolution of individual regions \citep[see][]{KAWAMURA09,SCHRUBA10,KRUIJSSEN14}, and so to better access the time-averaged behavior of the ISM. The evolutionary effects revealed at high resolution are explored in E.~Schinnerer, A.~Hughes et al. (in preparation) and M.~Chevance, J.~M.~D. Kruijssen et al. (in preparation).

In practice, we record the properties of M51 at each point in a hexagonally-packed, half beam-spaced grid \citep[see][]{LEROY13}. Figure~\ref{fig:sampling} shows the individual sampling points for these two grids, overlaid on the {\em Herschel} $70\mu$m map \citep{MENTUCHCOOPER12}.

\subsection{Data}

\citet{SCHINNERER13} and \citet{PETY13} present PAWS, which mapped CO~\mbox{(1-0)} emission from the central region of M51 at $1.16\arcsec \times 0.97\arcsec \sim 1.06\arcsec \sim 40$~pc resolution with ${\sim} 5$~km~s$^{-1}$ velocity resolution. PAWS includes short and zero-spacing information. \citet{SCHINNERER13} also summarize the multiwavelength data available for M51, with references (see their Table 2).

We also use broad band maps of IR emission from {\em Herschel} and {\em Spitzer}. These were obtained as part of the {\em Spitzer} Infrared Nearby Galaxy Survey \citep[][]{KENNICUTT03} and the {\em Herschel} Very Nearby Galaxies Survey \citep[][]{MENTUCHCOOPER12}.

\subsection{Measurements}
\label{sec:meas}

{\em Integrated CO Intensity:} At $30\arcsec$ resolution, we use the PAWS single dish map \citep{PETY13} to measure the integrated CO intensity. At $10\arcsec$, we convolve the combined interferometer and single dish cube to a coarser $10\arcsec$ resolution and measure the integrated intensity from this degraded map. As discussed by \citet{PETY13}, the deconvolution of the hybrid 30m+PdBI map recovers 99\% flux of the galaxy observed with the IRAM \mbox{30m}.

To collapse the $30\arcsec$ and $10\arcsec$ cubes to integrated intensity measurements, we sum over a broad velocity window from $-70$ to $+70$~km~s$^{-1}$ about the local mean velocity. The signal-to-noise in CO~\mbox{(1-0)} is very high, so some empty bandwidth is not a concern. We estimate the associated uncertainty by measuring the rms noise of the convolved line cube from the signal-free region. Then the statistical uncertainty in the integrated intensity is the sum in quadrature of the per-channel intensity noise across all channels in the velocity integration window multiplied by the velocity width of a channel.

{\em Total Infrared Surface Brightness:} We convolve the IR data to have Gaussian beams using the kernels of \citet{ANIANO11}. Then, using the formulae of \citet{GALAMETZ13}, we combine {\em Spitzer} 24 $\mu$m and 70 $\mu$m intensities with {\em Herschel} 160 $\mu$m and 250 $\mu$m intensities to estimate a total infrared luminosity (TIR) surface brightness, $\Sigma_{\rm TIR}$, for each resolution element. This is our basic measure of star formation activity throughout this paper.

At $10\arcsec$ resolution, we can only use the {\em Herschel} 70 $\mu$m data. We calculate the coefficient to translate $I_{70}$ to $\Sigma_{\rm TIR}$ by comparing the two quantities at $30\arcsec$ resolution, where we know $\Sigma_{\rm TIR}$ from the four-band calculation following \citet{GALAMETZ13}. In the PAWS field, the ratio $\Sigma_{\rm TIR} / I_{70}$ varies modestly as a function of radius, presumably reflecting a radial change in dust temperature. We find:

\begin{eqnarray}
\label{eq:70totir}
\frac{\Sigma_{\rm TIR}}{I_{70}} &=& 
10^6 \begin{cases}
f(r_{\rm gal}) {\rm~if~} r_{\rm gal} < 2.5~{\rm kpc} \\
2.96 {\rm~if~} r_{\rm gal} > 2.5~{\rm kpc} \\
\end{cases} \\
\nonumber {\rm where}\quad f(r) &=& 1.93 + 0.01r +0.28r^2 - 0.048r^3~.
\end{eqnarray}

\noindent Here $r_{\rm gal}$ refers to the deprojected galactocentric radius and $f(r)$ is a polynomial fit to the ratio $\Sigma_{\rm TIR}/I_{70}$ as a function of $r$ inside $r_{\rm gal} = 2.5$~kpc. $\Sigma_{\rm TIR}$ has units of L$_\odot$~kpc$^{-2}$, $I_{70}$ has units of MJy~sr$^{-1}$, and $r_{\rm gal}$ has units of kpc. Outside $r_{\rm gal} \sim 2.5$~kpc, the ratio appears flat. The appendix compares SFRs derived from $I_{70}$ using this approach to those using $\Sigma_{\rm TIR}$ at $\theta = 30\arcsec$. The two show a median ratio of $1$, less than 10\% scatter and no clear systematics across the PAWS field. 

At $30\arcsec$, we measure the rms scatter in $\Sigma_{\rm TIR}$ from the low intensity regions of the map to be ${\sim} 2.5 \times 10^6$ L$_\odot$~kpc$^{-2}$. At $10\arcsec$, using only the $70\mu$m data, the rms scatter is higher, ${\sim} 8.5 \times 10^6$ L$_\odot$~kpc$^{-2}$.

{\em Cloud Scale Properties:} We measure the intensity-weighted cloud-scale properties of the gas in each beam following \citet{LEROY16}. In brief, we begin with the native $40$~pc resolution PAWS cube. We recenter each spectrum about the local mean velocity. Next, we weight each spectrum by the integrated intensity along the line-of-sight and convolve from $1\arcsec \approx 40$~pc resolution (our ``measurement scale'') to $10\arcsec \approx 370$~pc or $30\arcsec \approx 1.1$~kpc (our ``averaging scales''). From these intensity-weighted, stacked spectra, we measure the integrated intensity and line width of the gas. This cross-scale weighted averaging also resembles that by \citet{OSSENKOPF02}.

Because of the intensity ($\sim$ mass for fixed $\alpha_{\rm CO}$) weighting, this approach captures the high resolution structure of the emission within each larger averaging beam. \citet{LEROY16} demonstrated that the results match those from mass-weighted averages of cloud catalogs well, but with far fewer assumptions. We write the resulting measurements as, e.g., $\sdavg$. This is read as ``the mass-weighted average $40$~pc surface density within a larger beam\footnote{More rigorously, following \citet{LEROY16} we would also indicate the size of that larger beam (the ``averaging scale'') when quoting $\sdavg$. In this paper the plots, discussion and tables make it clear whether $\sdavg$ refers to an averaging scale of $370$~pc ($\sdavg_{\rm 370pc}$ or $1.1$~kpc ($\sdavg_{\rm 1.1kpc}$.}.'' We focus on three such measurements: 

\begin{enumerate}[leftmargin=*]
\item The {\em cloud-scale molecular gas surface density, $\sdavg$}. This is a linear translation of the integrated intensity, $\sdavg = \alpha_{\rm CO}~\left< I_{\rm 40pc} \right>$, where $\alpha_{\rm CO}$ is our adopted CO-to-H$_2$ conversion factor. If the line-of-sight length of the gas distribution, $h$, is known or assumed, then $\sdavg$ can be used to estimate the volume density of the gas on $40~{\rm pc} \times h$ scales, $\rhosdavg$. From this, one can estimate the gravitational free-fall time, $\tffavg$. We show in Section \ref{sec:clouds} that for published Milky Way and M51 cloud catalogs, $\Sigma_{\rm mol}$ and $\rho$ do correlate well.

\item The {\em rms line width of CO, $\sigavg$}, measured from the ``equivalent width'' and corrected for channelization and channel-to-channel correlation following \citet{LEROY16}. For a given temperature and when the line width is purely turbulent in nature, this corresponds to the turbulent Mach number, $\mathcal{M}$. $\sigavg$ may also contain a contribution from bulk motions unresolved at the $40$~pc resolution of PAWS \citep[][and S.~Meidt et al., submitted]{COLOMBO14B,MEIDT13}. Thermal contributions to the line width are expected to be small.

\item The {\em dynamical state of the gas, as traced by the ratio $\bavg$ with $b \equiv \Sigma_{\rm mol} / \sigma^2$.} This ``boundedness parameter'' also relies on an adopted CO-to-H$_2$ conversion factor. Within a length scale $b$ is proportional to UE/KE, the ratio of potential energy (UE) to kinetic energy (KE). This is the inverse of the virial parameter, $b^{-1} \propto \alpha_{\rm vir} \approx 2 {\rm KE}/{\rm UE}$. When $b$ is high, the gas should be more gravitationally bound.

Also, within a length scale $\bavg \propto \tau_{\rm cross}^2 / \tau_{\rm ff}^2$, where $\tau_{\rm ff} \propto 1 / \sqrt{M/R^3}$ is the free-fall time and $\tau_{\rm cross} \sim R / \sigma$ is the crossing time for the measured velocity dispersion. This ratio has been highlighted by \citet{PADOAN12} as a key driver for the star formation efficiency per free fall time.

Note that in this paper we focus on $b \equiv \Sigma_{\rm mol} / \sigma^2$. This differs from  the $B \equiv I_{\rm CO}/\sigma^2$ discussed in \citet{LEROY16} by a factor of $\alpha_{\rm CO}$, so that $b = \alpha_{\rm CO} B$. While $B$ has the advantage of being directly computed from observable quantities, $b \propto \alpha_{\rm vir}^{-1}$ is more closely linked to the physical state of the gas.
\end{enumerate}

\noindent We estimate uncertainties in $\sdavg$, $\sigavg$, and $\bavg$ using a Monte Carlo approach. We measure the noise in the stacked, shuffled intensity weighted spectra from the signal-free region. Then we realize 100 versions of each spectrum, adding random noise to the real spectrum. For each case, we remeasure $\sdavg$, $\sigavg$, and $\bavg$. We compare these to our measurements without added noise, which we take to be the true values for purpose of this exercise. The rms offset between the simulated noisy data and the true value yields our estimate of the noise. This approach is {\em ad hoc} but yields realistic statistical uncertainties and captures the covariance among the uncertainties on $\sdavg$, $\sigavg$, and $\bavg$.

\begin{figure*}
\plottwo{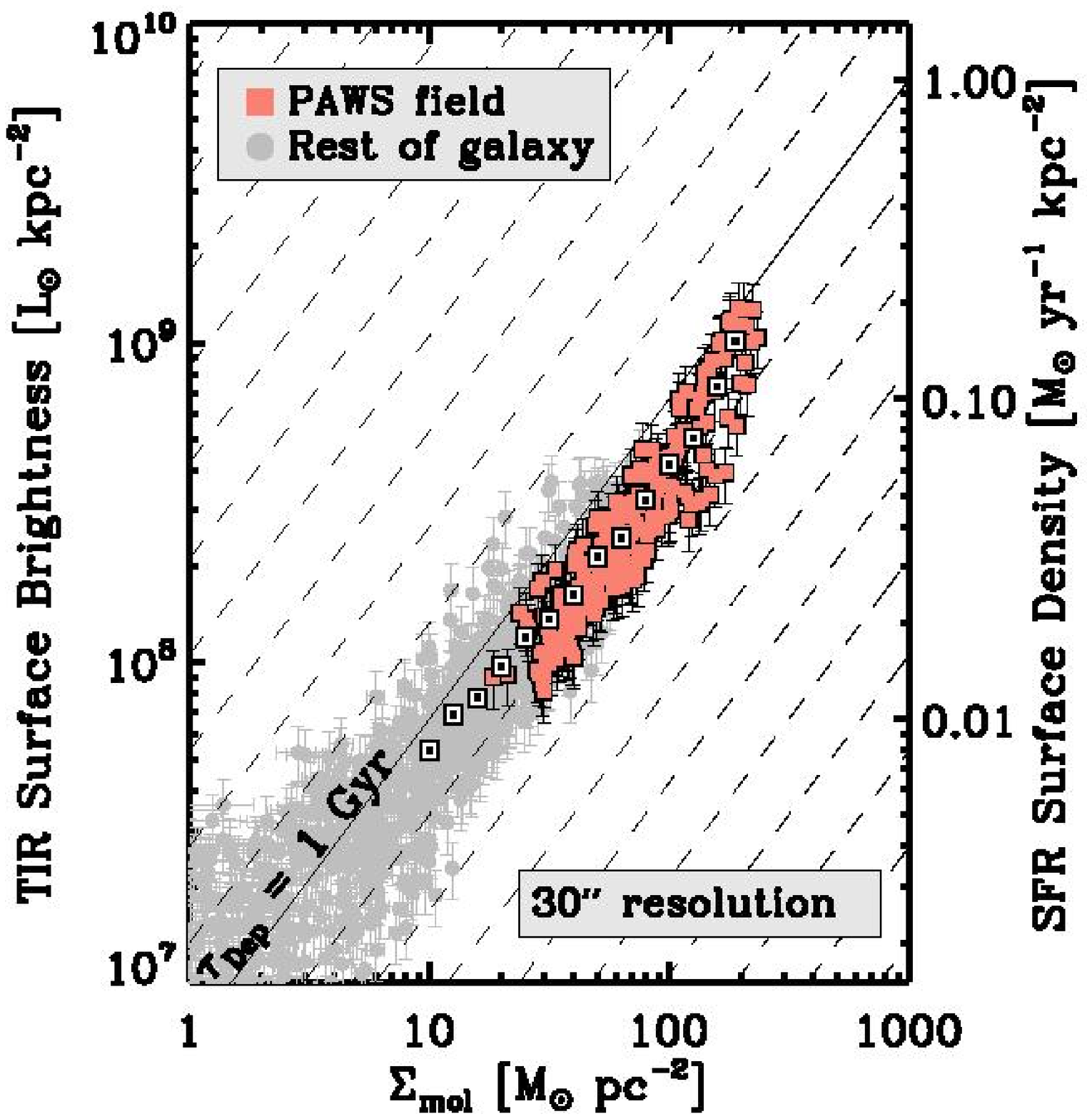}{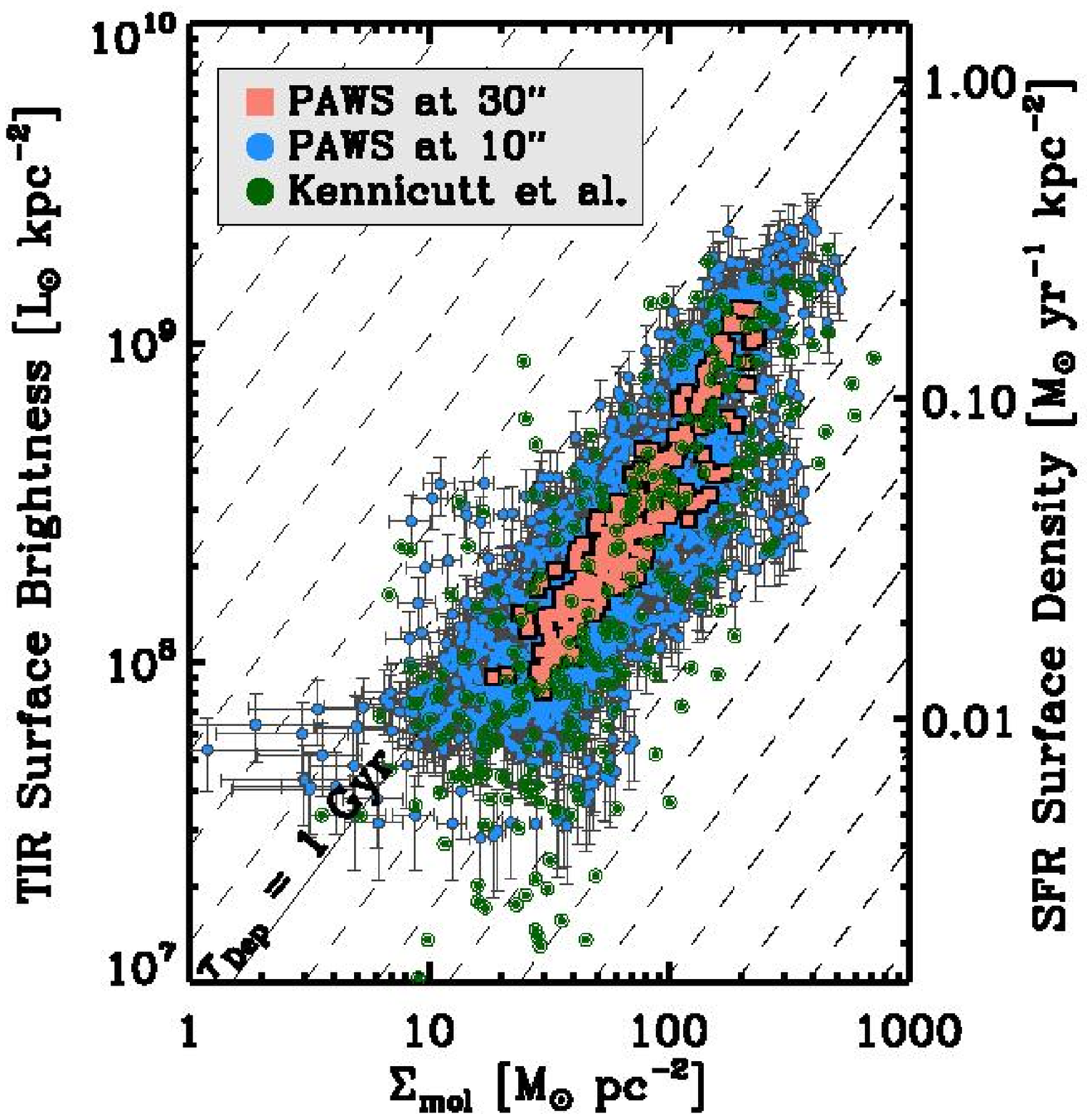}
\caption{IR-CO scaling relations in M51 over ({\em left}) the whole galaxy and ({\em right}) in the PAWS field. Diagonal lines indicate a fixed molecular gas depletion times (i.e., a fixed CO-to-IR ratio), spaced by factors of $2$. Red points in the left panel come from the PAWS field, which we study in this paper. Gray points show the measurements from the PAWS single dish map outside the interferometric survey area. Error bars show statistical uncertainty on individual points. In the left panel square points show a running mean $\Sigma_{\rm TIR}$ as a function of $\Sigma_{\rm mol}$. The weak bowed shape visible in these point may explain some of the discrepancies in the star formation scaling law literature for M51, as different studies focused on different parts of the galaxy. The right panel shows results at both resolutions in the PAWS field and overplots the measurements from \citet{KENNICUTT07}, which target $13\arcsec$ apertures on star-forming peaks over a similar area. More variation in the CO-to-IR ratio is visible at high resolution, including both high and low $\tau_{\rm Dep}^{\rm mol}$ at high $\Sigma_{\rm mol} \gtrsim 100$~M$_\odot$~pc$^{-2}$.}
\label{fig:scaling_law}
\end{figure*}

\subsection{Conversion to Physical Parameters}
\label{sec:convtophys}

We report our results in terms of simple transformations of observable quantities into physical parameters.

{\em Galactocentric Coordinates:} Following \citet{SCHINNERER13}, we assume an inclination $i=22\arcdeg$ \citep{COLOMBO14B} and a position angle ${\rm P.A.} = 172\arcdeg$ \citep{COLOMBO14B} with the galaxy center at $\alpha_{2000} = 13^{\rm h}~29^{\rm m}~52.7^{\rm s}$, $\delta_{2000} = +47\arcdeg~11\arcmin~43\arcsec$ \citep{HAGIWARA07}. We adopt the $7.6$~Mpc distance of \citet{FELDMEIER97} and \citet{CIARDULLO02}.

{\em CO~\mbox{(1-0)}-to-H$_2$:} We estimate H$_2$ mass from CO~\mbox{(1-0)} emission using a CO-to-H$_2$ conversion factor $\alpha_{\rm CO} = 4.35$~\acounits , which includes helium. This is a standard value for the Galaxy \citep{BOLATTO13A}. In the appendix, we show that a dust-based approach following \citet{SANDSTROM13} and \citet{LEROY11} suggests approximately this value. \citet{SCHINNERER10} came to the same conclusion via a multi-line CO analysis of the spiral arms. B.~Groves et al. (in preparation) show that this value applies with only weak variations across the disk of M51a using several independent methods. 

There have been other values suggested for M51, mostly lower than Galactic by a factor of $\approx 2$ based on dust observations \citep[e.g. ][though see our appendix]{NAKAI95,WALL16}. We discuss the impact of a lower conversion factor in the text.

{\em $\Sigma_{\rm TIR}$ to $\Sigma_{\rm SFR}$:} When relevant, we recast the TIR surface brightness as an SFR surface density using the conversion of \citet{MURPHY11}, which assumes a \citet{KROUPA01} initial mass function and reduces to 

\begin{equation}
\label{eq:sfr}
\frac{\Sigma_{\rm SFR}}{{\rm M}_\odot~{\rm yr}^{-1}~{\rm kpc}^{-2}} \approx 1.5 \times 10^{-10}~\frac{\Sigma_{\rm TIR}}{{\rm L}_\odot~{\rm kpc}^{-2}}~.
\end{equation}

A large body of work explores the subtleties of SFR estimation, often in M51 \citep[e.g.,][]{CALZETTI05,LEROY08,BLANC09,LIU11,LEROY12}. Our focus in this paper is new diagnostics of the molecular medium. Given the overwhelming extinction in the inner region of M51, we adopt the simple, widely accepted SFR diagnostic of TIR surface brightness. As a check, the appendix shows the impact of several alternative SFR prescriptions on our inferred molecular gas depletion time at $30\arcsec$ resolution. These matter mainly to the overall normalization. By using the TIR emission, it is likely that we somewhat overestimate $\Sigma_{\rm SFR}$. One of our key findings is that $\effsdavg$ is low (Section \ref{sec:eff}); this result would be even stronger if we used a tracer that yields lower $\Sigma_{\rm SFR}$. The systematic trends appear weak and, when present, go opposite the sense needed to yield a fixed $\effsdavg$. We intend to revisit this assumption in more detail in future work, ideally using an extinction-robust SFR tracer with high angular resolution to measure SFR on the scale of individual clouds. 

\subsection{Mapping to Dynamical Region}
\label{sec:regionassign}

Galactic dynamics relate to molecular gas structure and star formation in M51 \citep[e.g., see][]{KODA09,HUGHES13A,COLOMBO14A,MEIDT15,SCHINNERER17}. With this in mind, we separate our correlation analysis by dynamical regions (Section \ref{sec:region}). To do this, we use the dynamical region masks created by \citet{COLOMBO14A}. We use their simplified region definition, which breaks the PAWS field into ``arm'', ``interarm'', and ``central'' regions. For each $10\arcsec$ or $30\arcsec$ sampling point, we convolve the PAWS integrated CO intensity map multiplied by the mask for each separate region to the working resolution. Then we note the fraction of the flux in each beam coming from each dynamical region. When most of the CO flux in a beam comes from one dynamical environment, we associate the results for that beam with that environment.

Note that the three-region version of the \citet{COLOMBO14A} mask may still group together physically distinct environments. We treat the upstream and downstream interarm regions together \citep[e.g., see][]{MEIDT15}, and the ``center'' groups together the star-forming central molecular ring and the nucleus, which is more quiescent and potentially contaminated by the active galactic nucleus \citep[AGN, e.g., see][]{QUEREJETA16}.

\section{Results}
\label{sec:results}

\begin{figure*}
\plotone{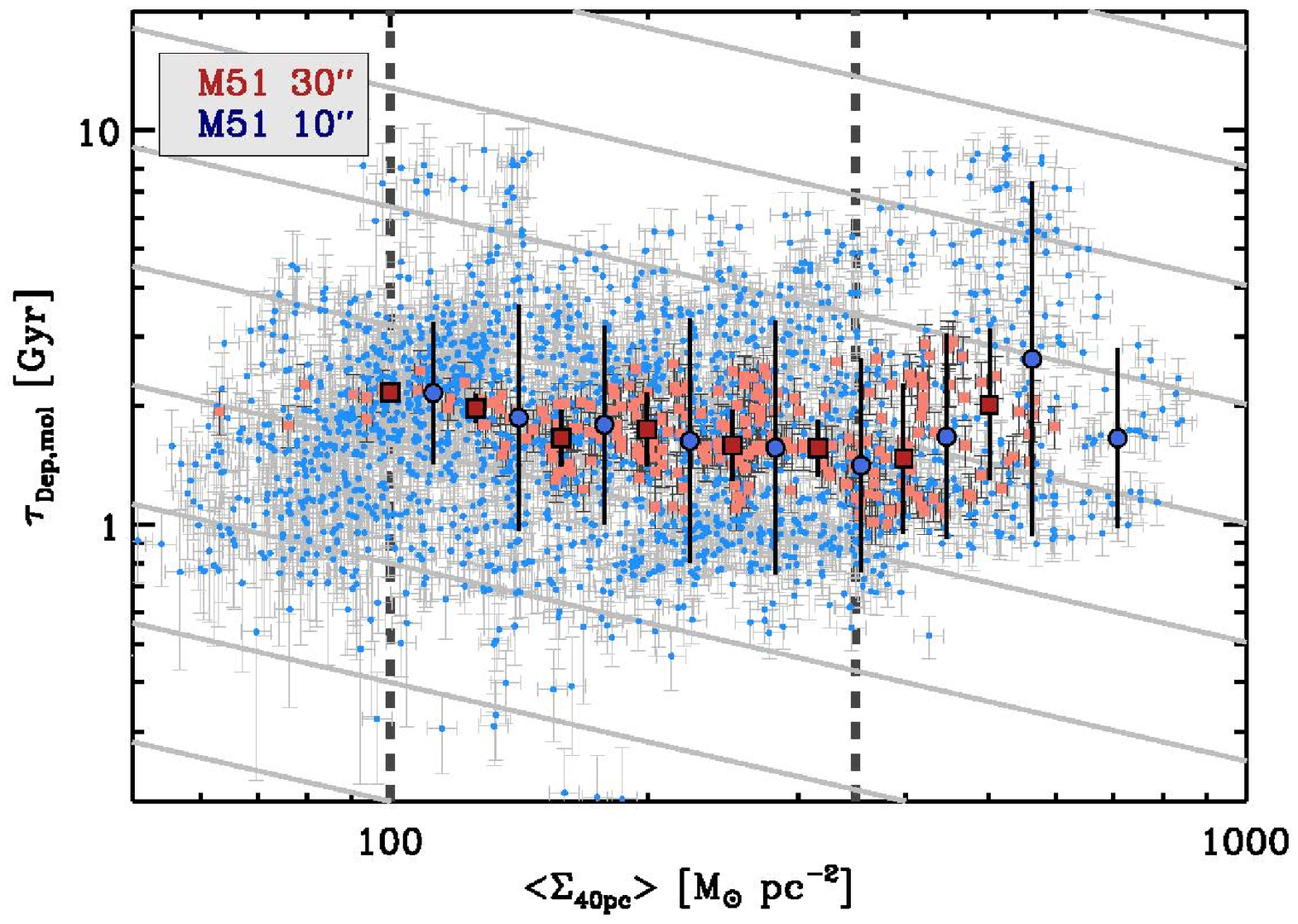}
\caption{Molecular gas depletion time, $\tau_{\rm Dep}^{\rm mol} \equiv \Sigma_{\rm mol} / \Sigma_{\rm SFR} \propto I_{\rm CO}/\Sigma_{\rm TIR}$, as a function of cloud-scale surface density, $\sdavg$. Note the difference from Figure \ref{fig:scaling_law}, which shows average surface density over large scales. Points here show mass-weighted average $40$~pc surface density with a $10\arcsec \approx 370$~pc (blue) and $30\arcsec \approx 1.1$~kpc (red) beam, and so reflect the local {\em cloud scale} surface density. Square points show median $\tau_{\rm Dep}^{\rm mol}$ in bins of $\sdavg$; error bars indicate the rms scatter in the bin. Gray lines show $\tau_{\rm Dep}^{\rm mol} \propto \sdavg^{-0.5}$, which is expected for a constant efficiency per free-fall time and $\rhoavg \propto \sdavg$. The dashed vertical lines indicate $\sdavg = 100$ and $350$~M$_\odot$~pc$^{-2}$. In this range we observe a mild anti-correlation between $\tau_{\rm Dep}^{\rm mol}$ and $\sdavg$, with denser gas forming stars at a higher normalized rate. Above $\sdavg = 350$~M$_\odot$~pc$^{-2}$, the sense of the correlation between $\sdavg$ and $\tau_{\rm Dep}^{\rm mol}$ shifts and higher surface density gas tends to form stars less effectively.}
\label{fig:tdep_sd}
\end{figure*}

Figure~\ref{fig:scaling_law} shows the scaling between TIR surface brightness, tracing $\Sigma_{\rm SFR}$, and CO intensity, tracing $\Sigma_{\rm mol}$. The left panel shows all of M51 at $30\arcsec \approx 1.1$~kpc resolution. The right panel includes only data from the PAWS field, plotting the $10\arcsec$ resolution measurements in blue, the $30\arcsec$ measurements in red, and the $13\arcsec$ measurements of \citet{KENNICUTT07} for selected apertures in green.

Over the whole of M51 (gray points), our data imply a molecular gas depletion time $\tau_{\rm Dep}^{\rm mol} \approx 1.5$~Gyr with ${\sim} 0.2$~dex scatter. In the PAWS field (red points), the numbers are about the same, $\tau_{\rm Dep}^{\rm mol} \approx 1.7$~Gyr with ${\sim} 0.1$~dex scatter. The median at $370$~pc resolution remains $\tau_{\rm Dep}^{\rm mol} = 1.6$~Gyr, but with larger scatter ${\sim} 0.3$~dex.

This resembles the $\tau_{\rm Dep}^{\rm mol} \approx 1{-}2$~Gyr found at the same resolution for a larger sample of similar nearby disk galaxies by \citet{LEROY13}. That study includes M51, but using different data. Our results agree qualitatively with their specific results for M51, including the presence of high IR-to-CO (low $\tau_{\rm Dep}^{\rm mol}$) regions in the inner galaxy. Here, the CO data have much higher signal-to-noise and we use only IR data to trace recent star formation. 

The right panel in Figure~\ref{fig:scaling_law} shows that our data also agree to first order with the measurements by \citet[][green, overlapping our blue points]{KENNICUTT07}. They targeted star-forming peaks with a different measurement strategy, $13\arcsec$ aperture photometry, and use yet another CO map \citep{HELFER03} and approach to $\Sigma_{\rm SFR}$, combining Paschen $\alpha$ and 24 $\mu$m emission.

Over the full area of M51 (left panel), the scaling between IR and CO exhibits a somewhat ``bowed'' shape moving from outside the PAWS field (the gray points at low $\Sigma_{\rm mol}$) to the inner disk (red points at high $\Sigma_{\rm mol}$). That is, the slope of the relation is slightly sublinear at low $\Sigma_{\rm mol}$ and superlinear at high $\Sigma_{\rm mol}$.

This curvature, which can be seen in the running mean (black-and-white squares) in the left panel, helps explain why different studies targeting M51 have come to apparently contradictory conclusions regarding the slope of the SFR-gas scaling relation \citep[e.g.,][]{LIU11,SHETTY13}. Those studying the inner part of the galaxy, especially at higher resolution using interferometers, see the superlinear slope evident at high surface densities. Those excluding the inner regions \citep{BIGIEL08,SHETTY13} and targeting a wider area find a modestly sub-linear slope. That is, given the curved shape of the relation in the left panel of Figure \ref{fig:scaling_law} we do not expect a single power law to fit all of M51. Note that this does not explain all of the scatter in the M51 literature, methodological differences including fitting and sampling strategy have also played a role \citep[e.g., see the appendix in][]{LEROY13}.

The right panel of Figure~\ref{fig:scaling_law} shows that at higher resolution, the IR surface brightness scatters more at fixed $\Sigma_{\rm mol}$, a result that has been measured before \citep{BLANC09,LEROY13}. The dependence of scatter on scale may be attributed to evolution of individual star-forming regions \citep[e.g.,][]{SCHRUBA10,KRUIJSSEN14}, and the $0.3$~dex scatter at ${\sim} 370$~pc resolution appears consistent with scatter expected from evolution in the \citet{KRUIJSSEN14} model. 

Our $370$~pc measurements may be more stochastic than the $1.1$~kpc calculations, but they also allow us to better isolate the physical conditions relevant to star formation. We capture more variation in local cloud populations and are better able to separate the galaxy into distinct regions. Below we find a larger range of ISM structure at $370$~pc than $1.1$~kpc, as well as stronger correlations between environment and ISM structure and distinct results for different dynamical regions. 

At $370$~pc resolution, we do observe substantial variation in $\Sigma_{\rm TIR}$ at a given $\Sigma_{\rm mol}$, including a wide range of $\Sigma_{\rm TIR}$ at high $\Sigma_{\rm mol} \gtrsim 100$~M$_\odot$~pc$^{-2}$. For $\Sigma_{\rm mol} = 30{-}100$~M$_\odot$~pc$^{-2}$, the median $\tau_{\rm Dep}^{\rm mol} \approx 2$~Gyr with $0.25$~dex scatter. For $\Sigma_{\rm mol} > 100$~M$_\odot$~pc$^{-2}$, the median $\tau_{\rm Dep}^{\rm mol}$ drops to $1.6$~Gyr but now with $0.37$~dex scatter. High IR-to-CO ratios (low $\tau_{\rm Dep}^{\rm mol}$) are preferentially found at high surface densities, which has helped fuel the result of superlinear power law scalings for $\Sigma_{\rm SFR}$ vs $\Sigma_{\rm mol}$ in M51 \citep{LIU11,MOMOSE13}. But there are also many lines of sight with high $\Sigma_{\rm mol}$ and relatively weak IR emission. These unexpected gas-rich, but relatively IR-weak, regions were highlighted by \citet{MEIDT13}. They argued that in these regions streaming motions suppress the collapse of gas.

Are these region-to-region variations in $\tau_{\rm Dep}^{\rm mol}$ driven by changes in the local structure of the gas? In the rest of this section, we explore this idea by comparing $\tau_{\rm Dep}^{\rm mol}$ to the local mean $40$~pc cloud scale surface density, velocity dispersion, and gravitational boundedness.

\subsection{Cloud Scale Surface Density and $\tau_{\rm Dep}^{\rm mol}$}
\label{sec:sd}

All other things being equal, high surface density gas should be denser, with a shorter collapse time, $\tau_{\rm ff}$. Do the variations in $\tau_{\rm Dep}^{\rm mol}$ in Figure \ref{fig:scaling_law} arise from changes in the cloud scale gas density across the galaxy? 

Figure~\ref{fig:tdep_sd} tests this expectation, plotting $\tau_{\rm Dep}^{\rm mol}$ as a function of $\sdavg$, the mass-weighted cloud scale surface density in each beam. Table~\ref{tab:corr} quantifies what we see in the Figure, reporting rank correlation coefficients between \sdavg\ and $\tau_{\rm Dep}^{\rm mol}$ for different ranges of \sdavg .

We do find a weak anti-correlation between $\tau_{\rm Dep}^{\rm mol}$ and $\sdavg$ over the range $\sdavg \approx 100{-}350$ M$_\odot$~pc$^{-2}$. Treating $\sdavg$ as the independent variable yields $\tau_{\rm Dep}^{\rm mol} \propto \sdavg^{-\alpha}$ with $\alpha = 0.25{-}0.35$ over this range. The rank correlation coefficient over this range is only $-0.14$, but still statistically significant.

Our simplest expectation would be $\tau_{\rm Dep}^{\rm mol} \propto \sdavg^{-0.5}$. This would be expected if $\rho \propto \sdavg$ (which appears reasonable, see Section \ref{sec:clouds}), and stars formed from gas with a fixed efficiency per $\tau_{\rm ff}$. Gray lines in the Figure illustrate this slope, which is steeper than the relation that we find. So  over $\sdavg \sim 100{-}350$~M$_\odot$~pc$^{-2}$ denser (or at least higher \sdavg ) gas does appear to form stars at a higher normalized rate, but the efficiency per free fall time decreases (weakly) as \sdavg\ increases.

At higher $\sdavg > 350$~M$_\odot$~pc$^{-2}$, $\tau_{\rm Dep}^{\rm mol}$ increases with increasing $\sdavg$, though with large scatter. This leads to the unexpected result pointed out by \citet{MEIDT13} that some of the least efficient star-forming regions in M51 have high cloud-scale molecular gas surface density. We show below that although these regions have high surface densities, they also appear to be less gravitationally bound (higher $\alpha_{\rm vir}$; Section \ref{sec:kin}). 

\begin{deluxetable*}{lcccccc}
\tabletypesize{\scriptsize}
\tablecaption{Rank Correlation Relating Cloud-Scale Structure with $\tau_{\rm Dep}^{\rm mol}$ and $\effsdavg$ \label{tab:corr}}
\tablewidth{0pt}
\tablehead{
\colhead{Quantity} & 
\colhead{vs. $\tau_{\rm Dep}^{\rm mol}$} &
\colhead{vs. $\tau_{\rm Dep}^{\rm mol}$} &
\colhead{vs. $\effsdavg$} &
\colhead{vs. $\effsdavg$} &
\colhead{vs. $\effsdavg$} &
\colhead{vs. $\effsdavg$} \\
\colhead{} &
\colhead{} &
\colhead{} &
\colhead{fixed $h$} &
\colhead{$h_{\rm dyn}$} &
\colhead{fixed $h$} &
\colhead{$h_{\rm dyn}$} \\
\colhead{} &
\colhead{at $\theta=30\arcsec$} &
\colhead{at $\theta=10\arcsec$} &
\colhead{at $\theta=30\arcsec$} &
\colhead{at $\theta=30\arcsec$} &
\colhead{at $\theta=10\arcsec$} &
\colhead{at $\theta=10\arcsec$} 
}
\startdata
\sdavg & & & & \\
$\ldots$ all data & $-0.14 (0.304)$ & $+ 0.02 (0.630)$ & $-0.59 (0.000)$ & $-0.78 (0.000)$ & $-0.47 (0.000)$ & $-0.61 (0.000)$ \\
$\ldots$ $100 < \sdavg < 350~\frac{{\rm M}_\odot}{{\rm pc}^{-2}}$ & $-0.17 (0.307)$ & $-0.14 (0.016)$ & $-0.51 (0.001)$ & $-0.67 (0.000)$ & $-0.15 (0.006)$ & $-0.30 (0.000)$ \\
$\ldots$ $\sdavg > 350~\frac{{\rm M}_\odot}{{\rm pc}^{-2}}$ & $+ 0.35 (0.202)$ & $+ 0.20 (0.089)$ & $-0.51 (0.039)$ & $-0.62 (0.010)$ & $-0.38 (0.002)$ & $-0.45 (0.000)$ \\
\sigavg & & & & \\
$\ldots$ all data & $+ 0.14 (0.278)$ & $+ 0.26 (0.000)$ & $-0.75 (0.000)$ & $-0.78 (0.000)$ & $-0.61 (0.000)$ & $-0.60 (0.000)$ \\
$\ldots$ $100 < \sdavg < 350~\frac{{\rm M}_\odot}{{\rm pc}^{-2}}$ & $+ 0.19 (0.254)$ & $+ 0.25 (0.000)$ & $-0.69 (0.000)$ & $-0.62 (0.000)$ & $-0.42 (0.000)$ & $-0.33 (0.000)$ \\
$\ldots$ $\sdavg > 350~\frac{{\rm M}_\odot}{{\rm pc}^{-2}}$ & $+ 0.84 (0.000)$ & $+ 0.64 (0.000)$ & $-0.89 (0.000)$ & $-0.84 (0.000)$ & $-0.72 (0.000)$ & $-0.66 (0.000)$ \\
\bavg & & & & \\
$\ldots$ all data & $-0.67 (0.000)$ & $-0.42 (0.000)$ & $+ 0.15 (0.270)$ & $-0.19 (0.145)$ & $+ 0.13 (0.004)$ & $-0.13 (0.005)$ \\
$\ldots$ $100 < \sdavg < 350~\frac{{\rm M}_\odot}{{\rm pc}^{-2}}$ & $-0.61 (0.001)$ & $-0.49 (0.000)$ & $+ 0.30 (0.064)$ & $-0.07 (0.672)$ & $+ 0.36 (0.000)$ & $+ 0.09 (0.124)$ \\
$\ldots$ $\sdavg > 350~\frac{{\rm M}_\odot}{{\rm pc}^{-2}}$ & $-0.79 (0.000)$ & $-0.64 (0.000)$ & $+ 0.73 (0.001)$ & $+ 0.56 (0.031)$ & $+ 0.59 (0.000)$ & $+ 0.46 (0.000)$ \\
\enddata
\tablecomments{Parenthetical values report the fraction of $1,000$ random re-pairings (accounting for an oversampling factor of $4$) that exceed the rank correlation of the true data. They can be read as Monte Carlo $p$ values. $\effsdavg$ with ``fixed $h$'' assumes a fixed line of sight depth of 100~pc. $\effsdavg$ with $h_{\rm dyn}$ uses Equation~\ref{eq:hdyn}.
\vspace{0.1in}}
\end{deluxetable*}

\subsection{Efficiency per Free-fall Time}
\label{sec:eff}

Given a distribution of gas along the line of sight, \sdavg\ traces \rhosdavg , the volume density of the gas averaged over the $\theta = 40$~pc beam. In turn, \rhosdavg\ determines the gravitational free-fall time, $\tau_{\rm ff}$. Contrasting $\tau_{\rm ff}$ with the measured $\tau_{\rm Dep}^{\rm mol}$ yields the efficiency per free fall time, $\epsilon_{\rm ff}$. An approximately fixed $\epsilon_{\rm ff}$ is argued to hold across scale and system by, e.g., \citet[][]{KRUMHOLZ12B,KRUMHOLZ05,KRUMHOLZ07}. More generally, $\tau_{\rm ff}$ is taken as the governing timescale for star formation, even when $\epsilon_{\rm ff}$ is low.

For gas with a depth $h$ along the line-of-sight,

\begin{eqnarray}
\label{eq:eff}
\rhosdavg &=& \left<\Sigma_{\rm 40pc}\right>~/~h \\
\nonumber \tffsdavg &=& \sqrt{3\pi / (32 G \rho )} = 81~{\rm Myr}~\left( \frac{\left<\Sigma_{\rm 40pc}\right>}{h_{\rm 100pc}} \right)^{-0.5} \\
\nonumber \effsdavg &=& \tffsdavg~/~\tau_{\rm Dep}^{\rm mol}
\end{eqnarray}

\noindent where $h_{100}$ is the depth of the molecular gas layer along the line-of-sight normalized to a fiducial value of $100$~pc. $\effsdavg$ is the efficiency per free-fall time, obtained by contrasting $\tau_{\rm Dep}^{\rm mol}$ with $\tffsdavg$.

\begin{figure*}
\plottwo{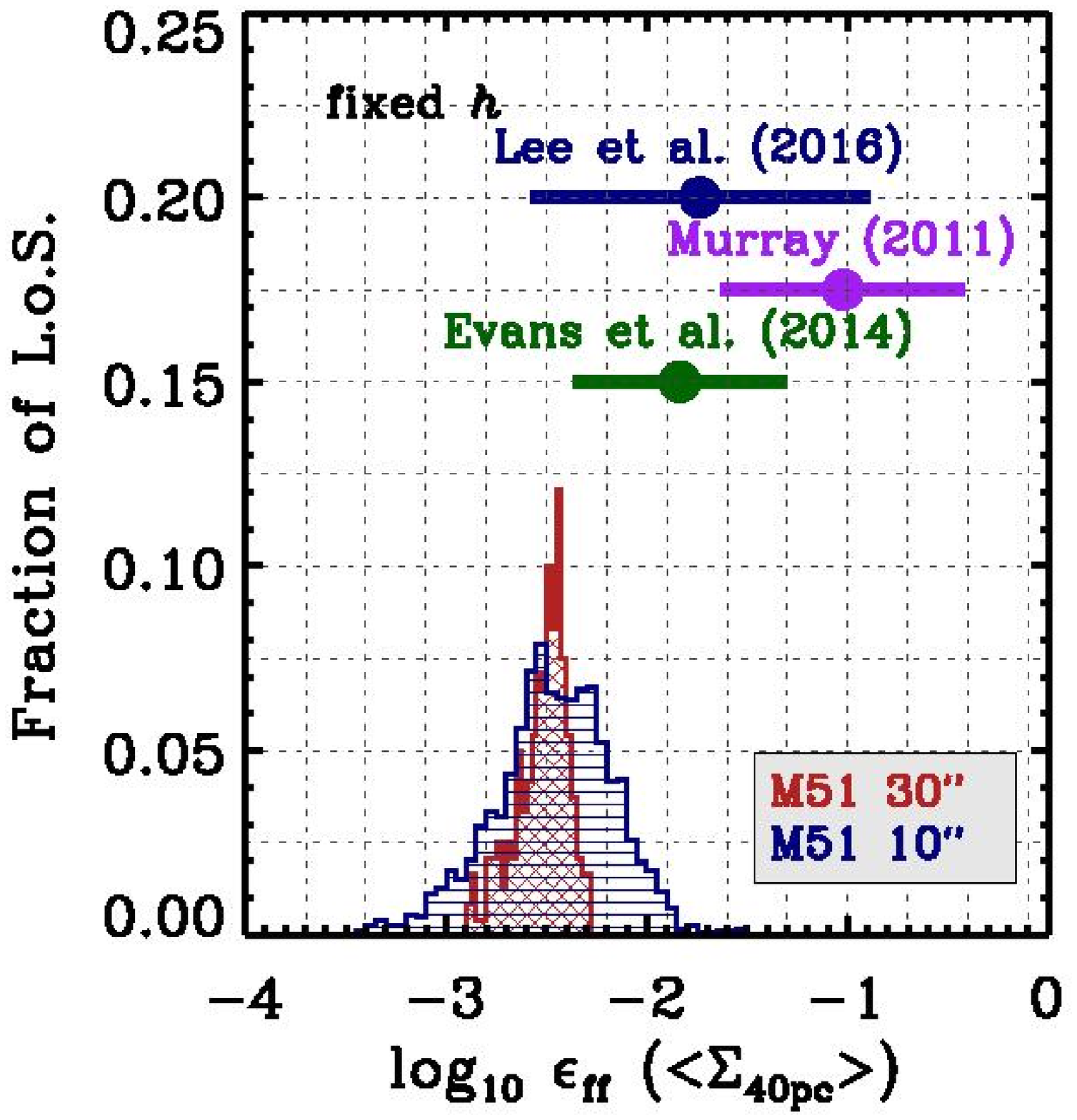}{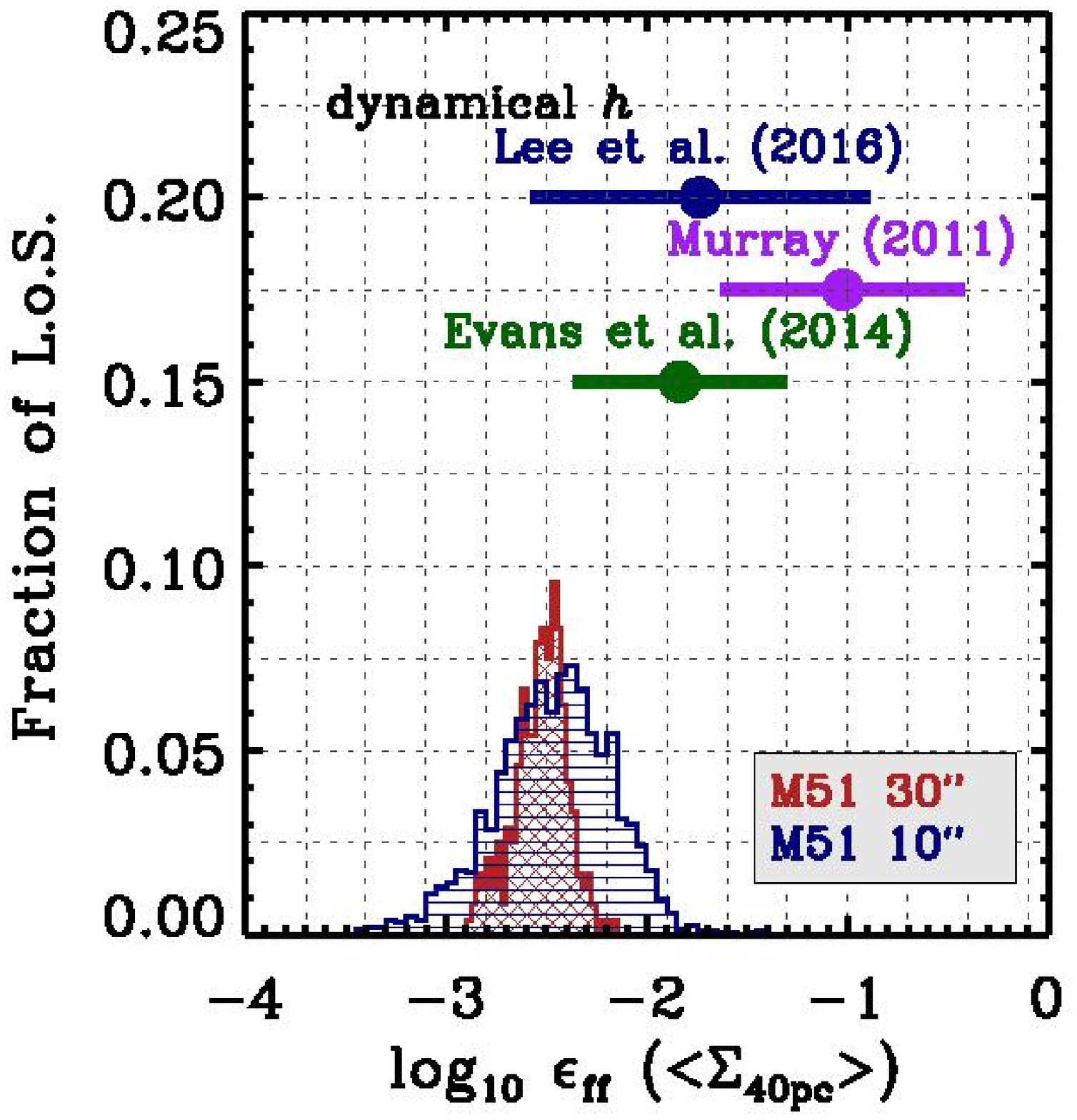}
\caption{Distributions of the implied efficiency of star formation per free-fall time, $\effsdavg$, ({\em left}) assuming a fixed line-of-sight depth of $100$~pc or ({\em right}) a variable depth $\propto b^{-1} = \sigma^2 / \Sigma$. Both resolutions yield low $\effavg \approx 0.003{-}0.0036$, with ${\sim} 0.3$~dex (${\sim} 0.1$~dex) scatter at $10\arcsec$ ($30\arcsec$) resolution. The scatter is similar between the two treatments of line of sight depth. Colored lines show the median and rms scatter from several Milky Way studies. We suggest that the differences with \citet{LEE16} and \citet{MURRAY11} reflect selection effects, and the difference with \citet{EVANS14} may reflect the influence of extended CO distributions around clouds. In both cases, more work is needed to resolve the discrepancy.}
\label{fig:eff}
\end{figure*}

The gray diagonal lines in Figure~\ref{fig:tdep_sd} show $\tau_{\rm Dep}^{\rm mol} \propto \Sigma_{\rm mol}^{-0.5}$. If $h_{100}$ remains fixed, then each of these lines corresponds to a fixed $\epsilon_{\rm ff}$. In Figure~\ref{fig:eff}, we show the distribution of \effsdavg\ implied by our measurements. We plot results for both working resolutions and show values for a fixed $h=100$~pc ({\em top}) and $h \propto \bavg ^{-1}$ ({\em bottom}, see explanation Section \ref{sec:los}). We also illustrate the range of $\epsilon_{\rm ff}$ measured by several Milky Way studies.

We find values of \effsdavg\ that are low in both the absolute sense and relative to theoretical and Milky Way values. We also find \effsdavg\ to vary as a function of environment and the local cloud population. Before discussing this in detail, we motivate our adopted $h$ (Section \ref{sec:los}) and demonstrate that $\sdavg$ indeed should be a good predictor of $\rhosdavg$ (Section \ref{sec:clouds}).

\subsubsection{What Line of Sight Depth to Use?}
\label{sec:los}

The depth of the gas layer along the line-of-sight $h$ affects \tffsdavg\ and so \effsdavg . We do not observe $h$, but we can make a reasonable estimate. The most common approach is to measure the radius of a GMC on the sky and then assume spherical symmetry. In cloud catalogs for the Milky Way \citep{HEYER09,MIVILLE17} and M51 \citep{COLOMBO14B}, most CO luminosity arises from clouds with radii ${\sim} 40{-}60$~pc. The left panel in Figure \ref{fig:clouds} shows the distribution of CO luminosity as a function of cloud radius for these three catalogs. The figure shows similar distributions for the \citet{COLOMBO14B} M51 catalog and the inner ($r_{\rm gal} < 8.5$~kpc) Milky Way portion of the recent \citet{MIVILLE17} catalog. In both cases, $68\%$ of the luminosity comes from clouds with $\sim 30~{\rm pc} < R < 95~{\rm pc}$, with the mid-point for CO emission $R \sim 60$~pc. The \citet{HEYER09} re-analysis of the \citet{SOLOMON87} Milky Way clouds (their ``A1'') yields slightly smaller cloud sizes, $\sim 20~{\rm pc} < R < 65~{\rm pc}$, with median $R \sim 40$~pc.

Estimates of the thickness of the molecular gas layer in both the Milky Way and M51 yield a similar value. \citet{HEYER15} compile estimates of the thickness of the molecular gas disk in the Milky Way (their Figure 6). They find $90{-}120$~pc (FWHM) within the Solar Circle. For M51, \citet{PETY13} assumed the molecular gas to be in hydrostatic equilibrium. Following \citet{KOYAMA09}, they calculated a mean FWHM thickness $\approx 94$~pc for the compact portion of the CO disk. If we consider the average density within FWHM $\approx 90$~pc, then the corresponding $h$ to use in Equations~\ref{eq:eff} is $h \approx 90 / 0.68 = 132$~pc.

Thus both estimates of the thickness of the molecular disk and GMC catalogs support our adopted $h \sim 100$~pc. Because $\tau_{\rm ff} \propto h^{-0.5}$, modest variations in $h$ do not have a large impact on $\effsdavg$. Still, we test the impact of varying $h$ by considering the case where the dynamical state of clouds (i.e., the virial parameter) is fixed. Then $M \propto r \sigma^2$ and $r \propto \sigma^2 / \Sigma = b^{-1}$. The same result applies for gas in a thin disk with only self-gravity. In this case:

\begin{equation}
\label{eq:hdyn}
h_{\rm dyn} \equiv 100~{\rm pc}~\bavg^{-1}~.
\end{equation}

Note that in this situation, where $b$ reflects a changing line of sight depth and not a changing dynamical state, $\bavg ^{-1}$ and $\sdavg$ are both linearly proportional to $\rhoavg$. Then we expect a similar relation of $\tau_{\rm Dep}^{\rm mol}$ to both variables. Below we show that this is not the case, and our best estimate is that $b$ in fact does reflect a changing dynamical state, not a changing line of sight depth. Thus, we consider the case of fixed $h=100$~pc to represent our basic result, and use Equation~\ref{eq:hdyn} to check the robustness of our conclusions.

\subsubsection{Cloud Surface and Volume Density}
\label{sec:clouds}

\begin{figure*}
\plottwo{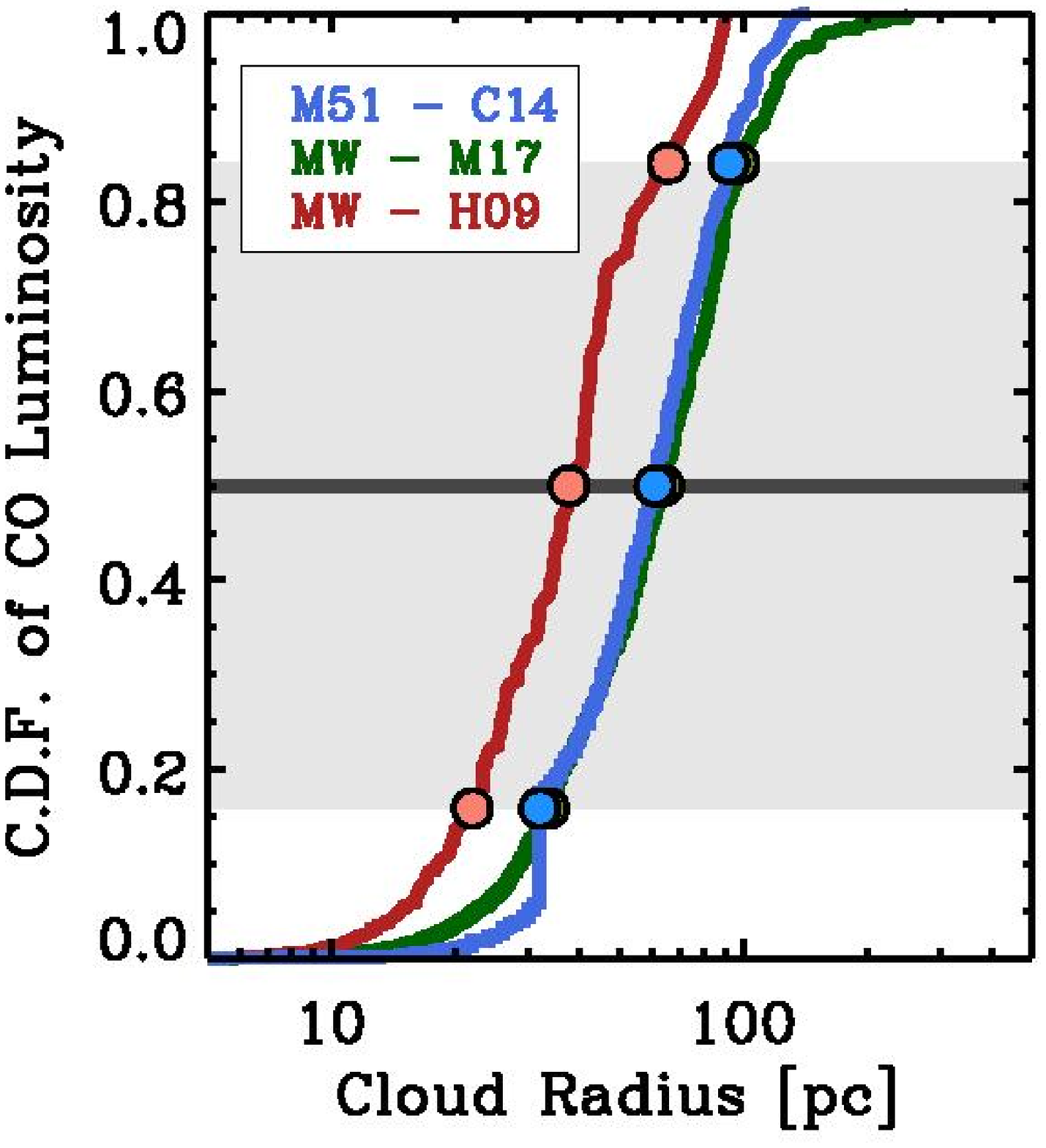}{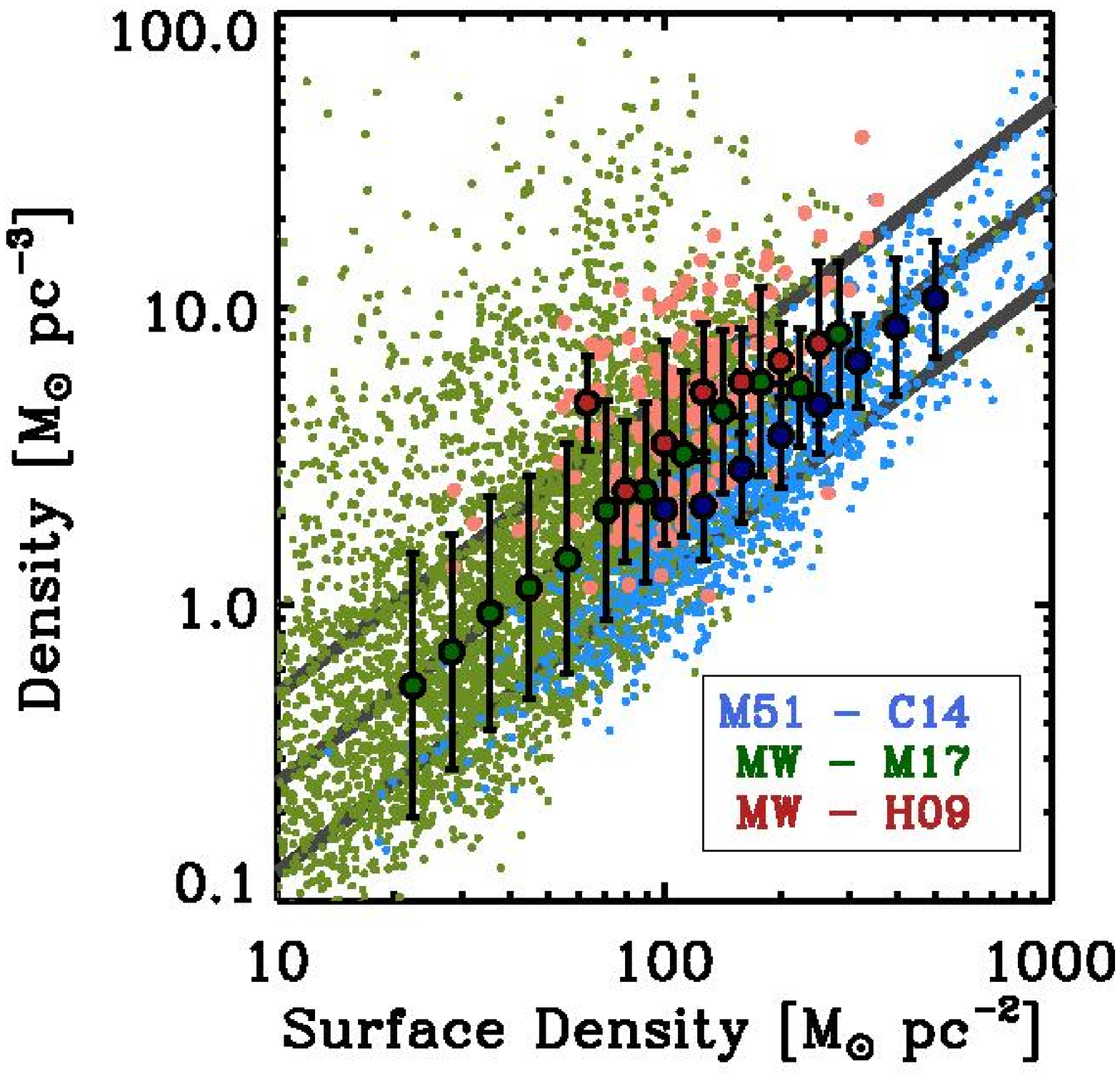}
\caption{Cloud radius, surface density, and volume density. ({\em  Left:}) The cumulative distribution of CO emission for the M51 GMC catalog of \citet[][C14]{COLOMBO14A} and two catalogs of Milky Way (MW) GMCs: the re-analysis of the \citet{SOLOMON87} clouds by \citet[][H09]{HEYER09} and inner galaxy $r_{\rm gal} < 8.5$~kpc clouds from the full-disk decomposition of \citet[][M17]{MIVILLE17}. Though there is some offset among the Milky Way measurements, most emission in the Milky Way and M51 catalogs comes from GMCs with $R \gtrsim 30$~pc and $R \lesssim 100$~pc. Along with estimates of the disk thickness in the Milky Way \cite[see][]{HEYER15} and M51 \citep[see][]{PETY13}, the plot motivates our fiducial line of sight depth $h=100$~pc for calculating the free-fall time. ({\em  Right:}) Density of molecular clouds $\rho = M / (4/3 \pi R^3)$ as a function of their surface density, $\Sigma = M / \pi R^2$, for the same three cloud catalogs. Surface density, our observable, correlates closely with volume density. This supports our treatment of the observable $\Sigma_{\rm 40pc}$ as a volume density diagnostic. The lines illustrate the relation for left-to-right fixed density $R=15,~30,~{\rm and}~60$~pc clouds.}
\label{fig:clouds}
\end{figure*}

The free fall time depends on the volume density, \rhoavg , but we observe the surface density, \sdavg . Although it has not been emphasized, these quantities do correlate well in current GMC catalogs. For the Milky Way and M51 catalogs mentioned above, the right panel in Figure \ref{fig:clouds} shows the volume density of each cloud, $\rho = M / (4/3 \pi R^3)$, as function of its surface density, $\Sigma = M / \pi R^2$. Surface and volume density correlate well, with rank correlation coefficients of $0.90$ \citep{COLOMBO14A}, $0.51$ \citep{HEYER09}, and $0.72$ \citep{MIVILLE17}.

Our mean inferred value for \effsdavg\ does not depend on the assumption that \sdavg\ maps perfectly to \rhoavg . Because $\tau_{\rm ff}$ depends weakly on $h$, it only matters that our adopted line-of-sight depth be roughly correct. But Figure \ref{fig:clouds} argues that a stronger case holds. The highly observable cloud scale surface density appears to be a reasonable proxy for the physically important, but hard to directly access volume density. More work on this topic is needed, but the right panel in Figure \ref{fig:clouds} offers an encouraging sign for extragalactic studies. Cloud scale mapping of CO surface brightness appears to offer a useful path to probe the mean volume density.

\subsubsection{Low Efficiency Per Free-fall Time}

In Figure~\ref{fig:eff}, \effsdavg\ varies between $10^{-3}$ and $10^{-2}$. For both treatments of $h$, the median \effsdavg\ is $3.6 \times 10^{-3}$ with $0.3$ dex scatter at $\theta = 10\arcsec$ resolution, and $3.0 \times 10^{-3}$ with $0.11$ dex scatter at $\theta = 30\arcsec$ resolution.

These values of \effsdavg\ are low in the absolute sense, with only $0.1{-}1\%$ of the gas converted to stars per collapse time. They are also low relative to some expectations from theory and previous work on the Milky Way, though they agree with previous indirect extragalactic estimates of $\effsdavg$.

{\em Comparison to Estimates at Large Scales:} Our $\effsdavg \approx 0.3\%$ agrees with the calculation by \citet{AGERTZ15}, who compared $\tau_{\rm ff}$ for Galactic GMCs to a typical $\tau_{\rm Dep}^{\rm mol}$ for nearby disk galaxies. In a similar vein, our median $\effsdavg$ is only a factor of $\sim 2$ lower than the estimate by \citet{MURRAY11} of a Milky Way disk-averaged $\epsilon_{\rm ff} \approx 0.6\%$.

Observations comparing dense gas, CO, and recent star formation also suggest a low $\epsilon_{\rm ff}$. \citet{GARCIABURILLO12} and \citet{USERO15} observed dense gas tracers, CO, and recent star formation in nearby star-forming galaxies. The combination of these three measurements is sensitive to the density of the gas and to the star formation per unit gas. Thus it depends on $\epsilon_{\rm ff}$, though in a model-dependent way. \citet{GARCIABURILLO12} and \citet{USERO15} argued that a low $\epsilon_{\rm ff} \approx 0.2\%$ appears to be required in order for their observations to match the turbulent cloud models of \citet[][]{KRUMHOLZ07}.

{\em Theoretical Values:} Our $\effsdavg \sim 0.3\%$ is lower than the $\epsilon_{\rm ff} \approx 1\%$ expected at the outer scale of turbulence by \citet{KRUMHOLZ05} and \citet{KRUMHOLZ12B}. Our values are about half of the $\epsilon_{\rm ff} \approx 0.5\%$ noted by \citet{MCKEE07}. They are also lower than the values commonly adopted by numerical simulations of galaxies \citep[e.g.][]{AGERTZ15} or found by simulations of individual star-forming regions \citep[e.g.,][]{PADOAN11}. 

We note that many of these predictions also depend on the virial parameter \citep[e.g.,][]{PADOAN02,KRUMHOLZ05}, with the Mach number, magnetic support, and type of turbulence also playing a role \citep[e.g.,][]{FEDERRATH12,FEDERRATH13}. In these cases, matching our observations may be primarily an issue of re-tuning these parameters, though some of these are also constrained by our data (see below).

{\em Comparison to Milky Way Results:} Our measured \effsdavg\ is significantly lower than the mean $\epsilon_{\rm ff} \approx 1.5\%$ found by \citet{EVANS14} for local clouds, and the median $\epsilon_{\rm ff} \approx 1.8\%$ found by \citet{LEE16} based on the \citet{MIVILLE17} Milky Way GMC catalog and WMAP-based SFRs. It is also much lower than the median $\epsilon_{\rm ff} \approx 9.5\%$ found by \citet{MURRAY11} for the GMCs associated with the brightest ${\sim} 32$ star-forming complexes in the Milky Way.

In the case of \citet{MURRAY11}, this discrepancy is expected. Those clouds were selected based on their association with active star formation, and may have among the highest SFR/M$_{\rm gas}$ in the Milky Way. Similarly, the cross matching of \citet{LEE16} recovers $\sim 80\%$ of the ionizing photon flux in their star forming complexes, but only $\sim 10\%$ of the GMC mass in the \citet{MIVILLE17} catalog. Our observations average over the entire life cycle of clouds present in a large averaging beam, and so can be expected to include the balance of GMC flux. While this has the advantage of better accessing the time averaged behavior of the gas, it also means that we cannot construct a measurement analogous to \citet{LEE16} and \citet{MURRAY11}. In the near future, with a $1\arcsec$ resolution extinction-robust SFR tracer, we would be able to associate individual clouds with star forming complexes, and so potentially access the same dynamical evolution of clouds that leads to the high $\epsilon_{\rm ff}$ in the \citet{LEE16} and \citet{MURRAY11} results.

Any similar bias towards only star forming clouds in the \citet{EVANS14} is less clear, but the discrepancy between our ``top down'' view and the local cloud measurements by \citet{EVANS14} has also been noted before \citep[see][]{HEIDERMAN10,LADA10,LADA12}. In detail, \citet{EVANS14} find a ${\sim} 5$ times shorter $\tau_{\rm Dep}^{\rm mol}$ for their clouds than we see for large parts of M51. They also find a ${\sim} 4$ times shorter $\tau_{\rm ff}$. One plausible explanation for the discrepancy is that \citet{EVANS14} focus on the part of a cloud with $A_V > 2$~mag ($\approx 20$~M$_\odot$~pc$^{-2}$). Including a massive extended envelope or diffuse component might bring both $\tau_{\rm Dep}^{\rm mol}$ and $\tau_{\rm ff}$ into closer agreement with our measured values.

\subsubsection{Possible Systematic Effects}

We argue that most of the discrepancy with Milky Way results can be understood in terms of scales sampled and selection effects. However, several systematic uncertainties could affect our measurement, include our star formation rate estimate, adopted CO-to-H$_2$ conversion factor, and line of sight depth.

{\em Star Formation Rate:} On average, we would need to be underestimating the SFR of M51 by a factor of 5 to bring our measurements into agreement with the local clouds of \citet{EVANS14}. Meanwhile, in the appendix we show many likely biases in $\Sigma_{\rm SFR}$ would render our TIR-based calculation an {\em over}estimate, including any IR cirrus term \citep{LIU11,LEROY12}. Note, however, that \citet{FAESI14} argue that there may be up to a factor of ${\sim} 2$ offset between the SFR estimates used in local clouds and the tracers used at larger scales, with the local measurements yielding higher values \citep[see also][]{LEWIS17}. This offset has the right sense, but would have to reach even larger magnitude to bring our observation into agreement with the local clouds. Also, note that \citet{LEE16} and \citet{MURRAY11} use ionizing photon rates, similar to extragalactic studies.

{\em CO-to-H$_2$ Conversion Factor:}  Our adopted $\alpha_{\rm CO}$ also affects \effsdavg . We adopt a Galactic conversion factor based \citet{SCHINNERER10}, \citet{COLOMBO14A}, the calculations in the appendix, and B.~Groves et al. (in prep.). Other work has claimed a lower conversion factor in M51 \citep[see][for a summary]{SCHINNERER10}. Although evidence from dust, multi-line analysis, and cloud virial masses support our assumption, the systematic uncertainties in any given determination remain substantial \citep[see][]{BOLATTO13B}. For a lower $\alpha_{\rm CO}$, we would derive a shorter $\tau_{\rm Dep}^{\rm mol}$, a longer $\tau_{\rm ff}$, and a higher \effsdavg , with $\effsdavg \propto \alpha_{\rm CO}^{-1.5}$. $\alpha_{\rm CO}$ has a stronger effect on \effsdavg\ because it affects both $\tau_{\rm ff}$ and $\tau_{\rm Dep}^{\rm mol}$. Therefore a conversion factor $0.5$ times Galactic would yield $\effsdavg \approx 0.85\%$.

{\em Line of Sight Depth:} The adopted line-of-sight depth, $h$, affects \effsdavg . As emphasized above, our adopted $h$ agrees with both cloud property estimates and modeling of the M51 gas disk. To increase our measured $\effsdavg$ from ${\sim} 0.3\%$ to ${\sim} 1\%$, we would need to {\em increase} $h$ by an order of magnitude, to ${\sim} 1$~kpc. Such a scale height disagrees with the measured cloud properties in M51. A more substantial uncertainty in this direction is the role of any ``diffuse'' CO disk. Up to 50\% of the CO emission in M51 has been argued to lie in an extended component \citep{PETY13}. The physical nature of such a component remains unclear, but in the limit that it has a large scale height and holds half the gas, $\effsdavg$ for the compact component could increase by a factor of $2$ to $\sim 0.6\%$. In fact, we do not expect this effect to be so strong, as the bright, compact structures in the combined PdBI+30m map do hold a large fraction of the flux \citep{LEROY16}, but 10s of percent of the CO might lie in such an extended phase. This topic certainly requires more investigation in both the Milky Way and other galaxies.

To summarize, our $\effsdavg \sim 0.3{-}0.36\%$ does represents our best estimate, though systematic uncertainties could plausibly raise this by a factor of $\sim 2$. Supporting this conclusion, we note that our calculation agrees within a factor of $2$ with previous large scale calculations. Because of the external perspective and averaging approach, we argue that our value represents the correct comparison point for any model aiming to predict a population-averaged $\effsdavg$. Cloud-by-cloud statistics will need to await future, high resolution SFR maps.

\subsection{Efficiency Per Free Fall Time, $\tau_{\rm Dep}^{\rm mol}$ \\ and Local Gas Properties}
\label{sec:eff_sd}

At $370$~pc resolution we find $0.3$~dex scatter in \effsdavg , and Figure \ref{fig:tdep_sd} shows a comparable scatter in $\tau_{\rm Dep}^{\rm mol}$. Beyond only estimating $\effsdavg$, we aim to understand how the mean gas properties in the beam and the region of the galaxy under consideration influence these two quantities. That is, how much of this scatter is random and how much results from changes in the local gas properties? Both $\tau_{\rm Dep}^{\rm mol}$ and $\effsdavg$ are of interest: $\tau_{\rm Dep}^{\rm mol}$ captures the SFR per unit gas, and represents our most basic observation metric of whether gas in a part of a galaxy is good or bad at forming stars. $\effsdavg$ captures the efficiency of star formation relative to direct collapse, with $\tau_{\rm ff}$ representing the most common reference point for current theoretical models.

\subsubsection{Surface Density}

\begin{figure*}
\plottwo{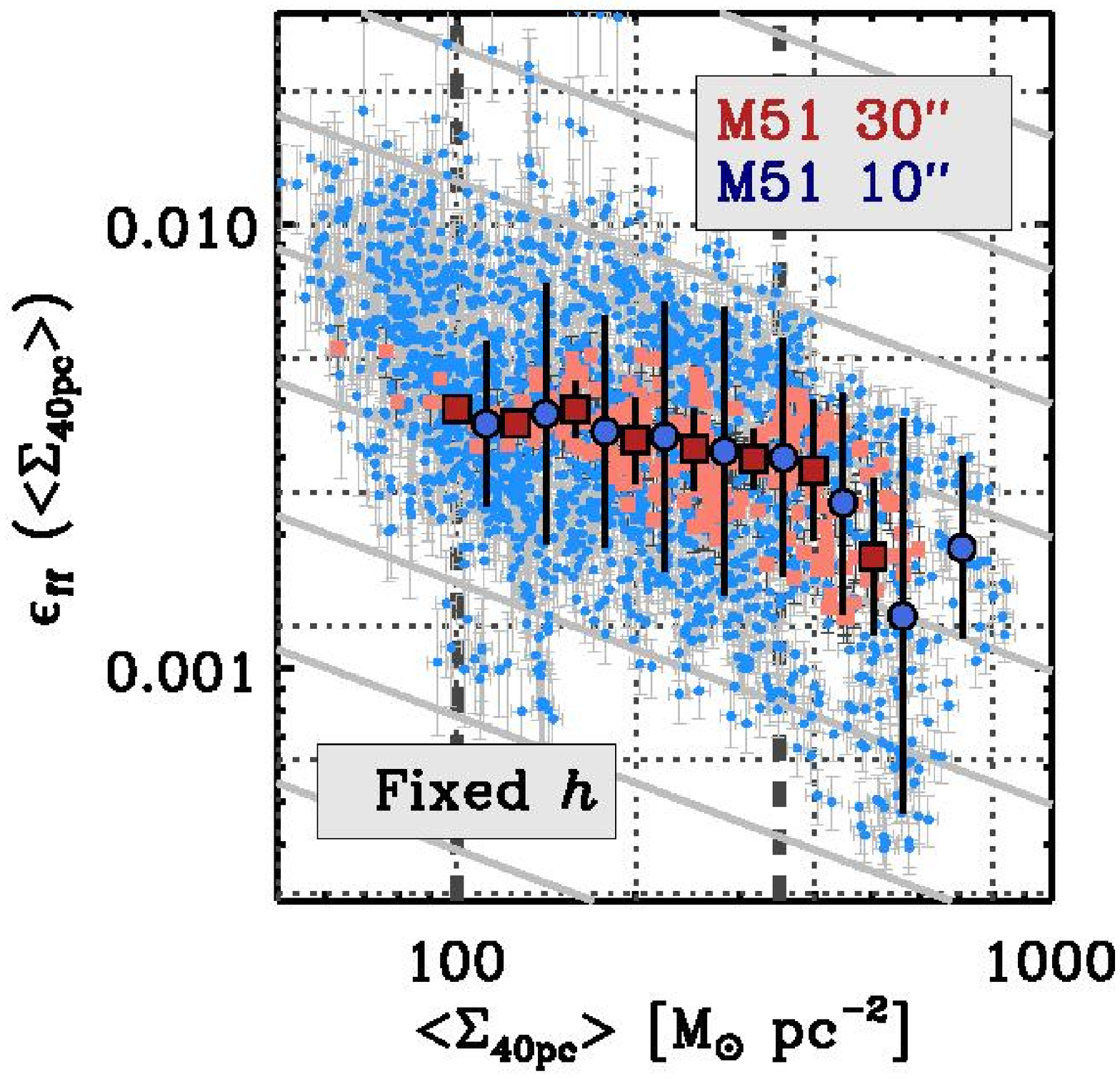}{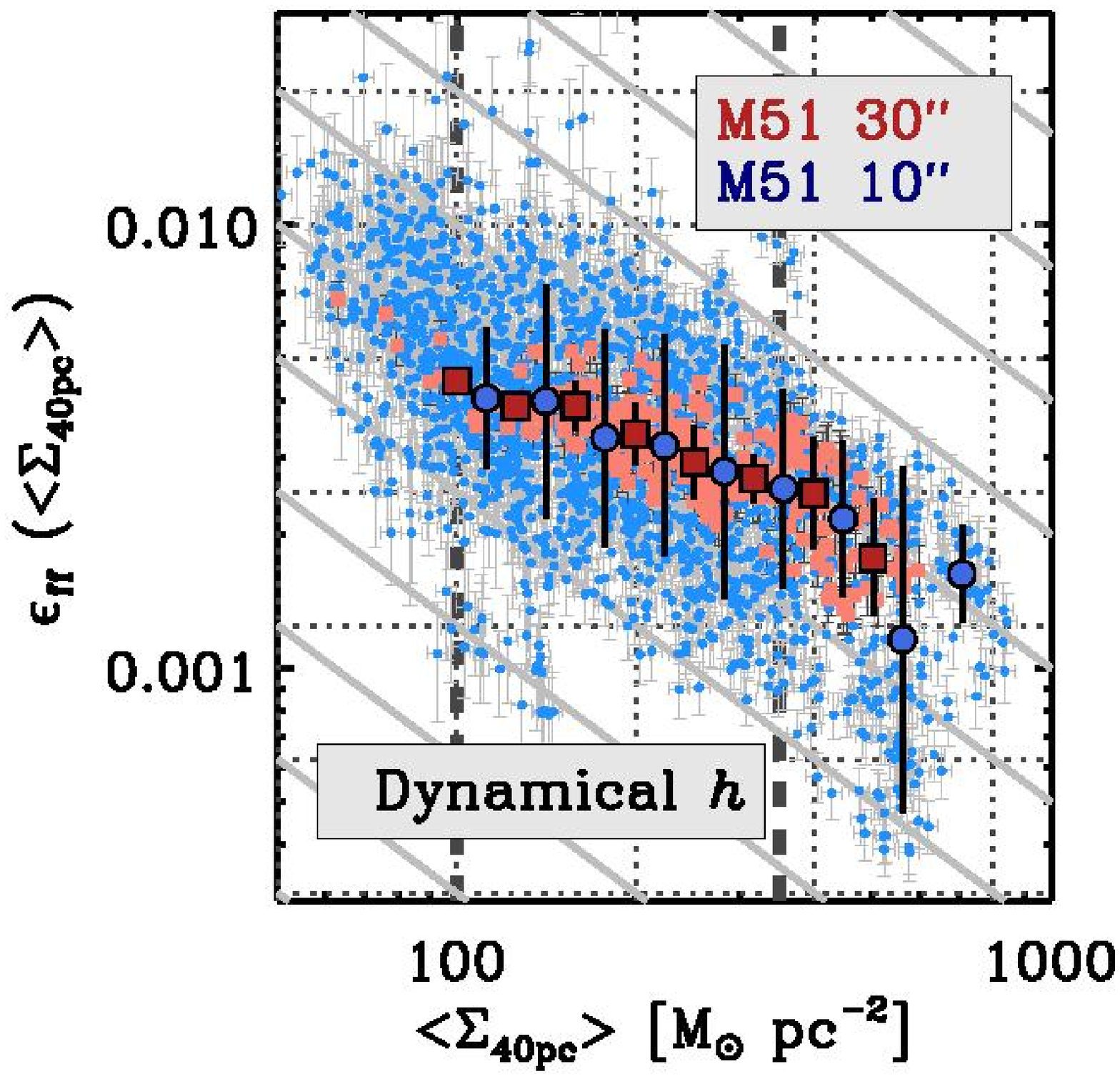}
\caption{\effsdavg\ as a function of \sdavg\ at $10\arcsec \approx 370$~pc (blue) and $30\arcsec \approx 1.1$~kpc (red) resolution for our two treatments of line of sight depth, ({\em left}) fixed $h$, our preferred approach and ({\em right}) a check using a dynamical estimate of the line of sight depth. Gray lines show the expectation if $\tau_{\rm Dep}^{\rm mol}$ does not correlate with $\tau_{\rm ff}$, which is  $\effsdavg \propto \sdavg^{-0.5}$ for a fixed $h$ and $\effsdavg \propto \sdavg^{-1}$ for $h \propto b^{-1} =  \sigma^2/\Sigma$. We observe a weak anticorrelation between \effsdavg\ and \sdavg\ at intermediate $\sdavg \approx 100{-}350$ M$_\odot$~pc$^{-2}$, and then a large drop in \effsdavg\ at higher $\sdavg > 350$~M$_\odot$~pc$^{-2}$.}
\label{fig:eff_sd}
\end{figure*}

Figure \ref{fig:tdep_sd} shows $\tau_{\rm Dep}^{\rm mol}$ as a function of cloud scale surface density. Figure \ref{fig:eff_sd} shows the corresponding plots for \effsdavg . As discussed above, $\tau_{\rm Dep}^{\rm mol}$ weakly anti-correlates with \sdavg\ over the range $\sdavg \approx 100{-}350$ M$_\odot$~pc$^{-2}$ and then increases, with large scatter towards higher densities. The observed $\sim -0.3$ slope relating $\tau_{\rm Dep}^{\rm mol}$ to \sdavg\ is shallower than that expected for a fixed \effsdavg . As a result, Figure \ref{fig:tdep_sd} shows $\effsdavg$ weakly decreasing with increasing $\sdavg$ for the fixed $h$ case. Though the slope in the right panel is shallow, Table~\ref{tab:corr} shows that \effsdavg\ does correlate with \sdavg\ over this range with good significance.

This trend in \effsdavg\ is weak compared to the large scatter until $\sdavg > 350$~M$_\odot$~pc$^{-2}$, at which point $\effsdavg$ drops precipitously. The high $\tau_{\rm Dep}^{\rm mol}$ at high \sdavg\ in Figure \ref{fig:tdep_sd} correspond to even lower \effsdavg . Thus the very high surface density parts of M51 \citep[the inner arms; see][and next section]{MEIDT13} are significantly less efficient than the rest of the galaxy at forming stars relative to the expectation for direct collapse ($\tau_{\rm ff}$). The most extreme values in Figure \ref{fig:eff_sd} reach $< 0.1\%$, though $\sim 0.2\%$ represents a more typical \effsdavg\ at these high \sdavg .

The right panel adopts our alternate treatment of $h$ (Equation \ref{eq:hdyn}). The main difference from the left panel is a stronger anti-correlation between \effsdavg\ and \sdavg\ at intermediate surface densities (see Tab~\ref{tab:corr}). The left panel represents our best estimate, but the consistency between the two suggests that our qualitative results are robust: there is some anti-correlation between \effsdavg\ and \sdavg\ at intermediate densities and even lower \effsdavg\ at high \sdavg .

Note that the axes in Figure~\ref{fig:eff_sd} are correlated because $\tau_{\rm ff} \propto \sdavg^{-0.5}$ in both panels. This built-in correlation is stronger in the right panel because for our dynamical scale height (Equation~\ref{eq:hdyn}) $h \propto \sdavg^{-1}$. The statistical uncertainty in \sdavg\ is small, ${\sim} 5\%$, and therefore we do not expect correlated noise to affect the results much. The larger issue is that if $\tau_{\rm Dep}^{\rm mol}$ and $\tau_{\rm ff}$ are unrelated, then $\effsdavg \propto \sdavg^{-0.5}$ for fixed $h$ by construction. That is, the null hypothesis that $\tau_{\rm ff}$ is not a governing timescale for star formation, we expect an anticorrelation in Figure \ref{fig:eff_sd}. This does not invalidate the measurement, but should be kept in mind when interpreting the plot.

\subsubsection{Velocity Dispersion}
\label{sec:kin}

\begin{figure*}
\epsscale{0.9}
\plottwo{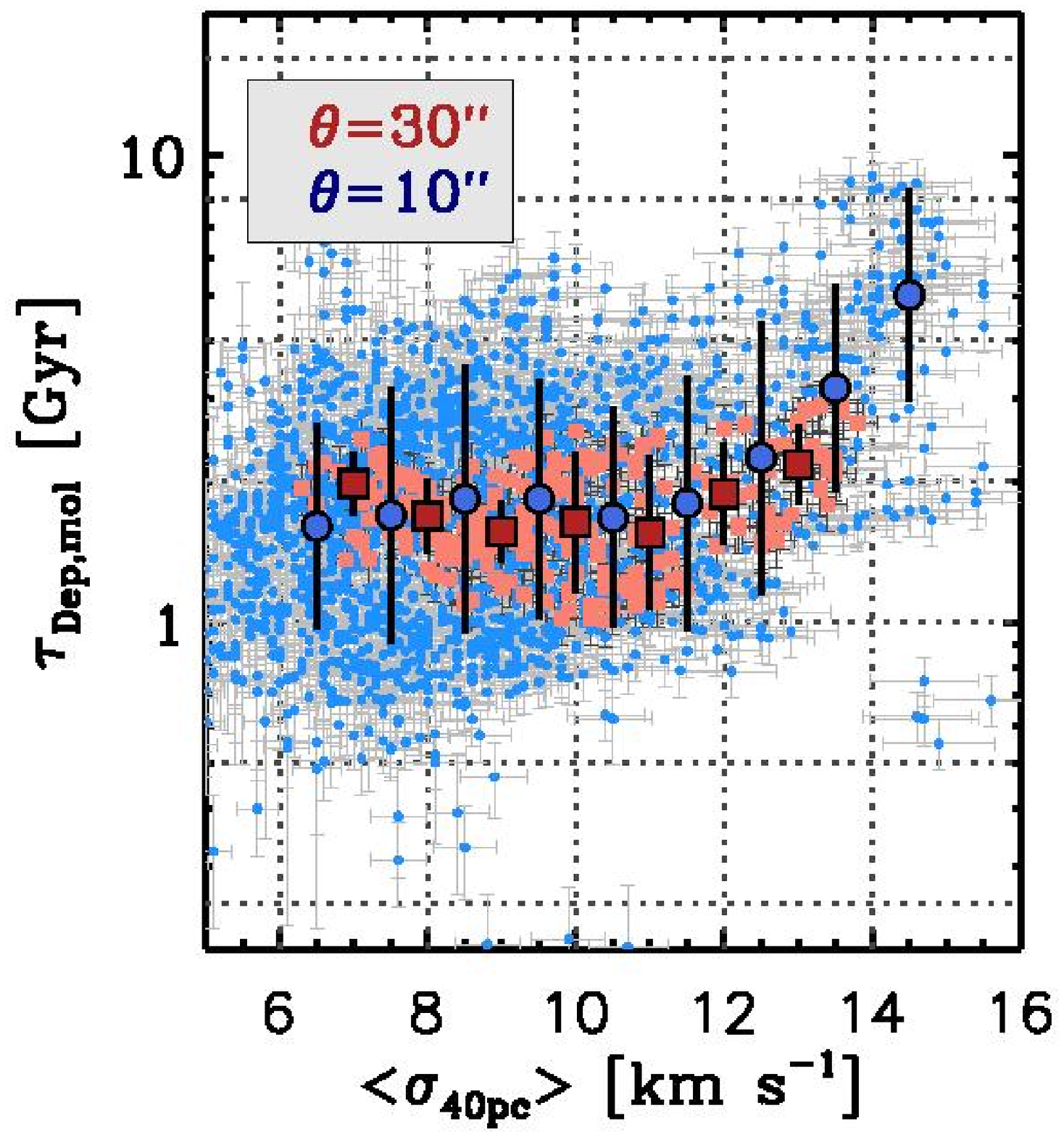}{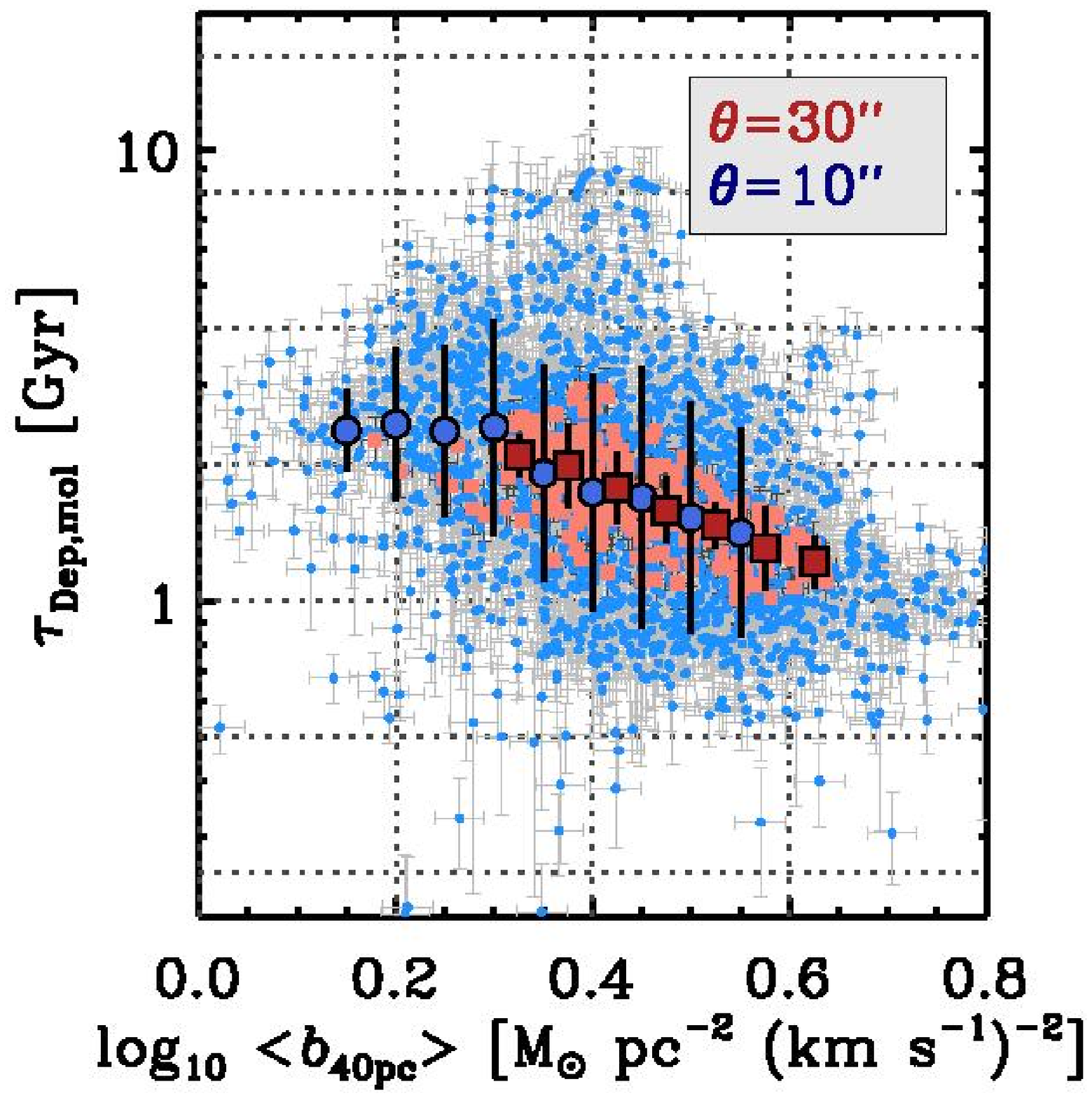}
\plottwo{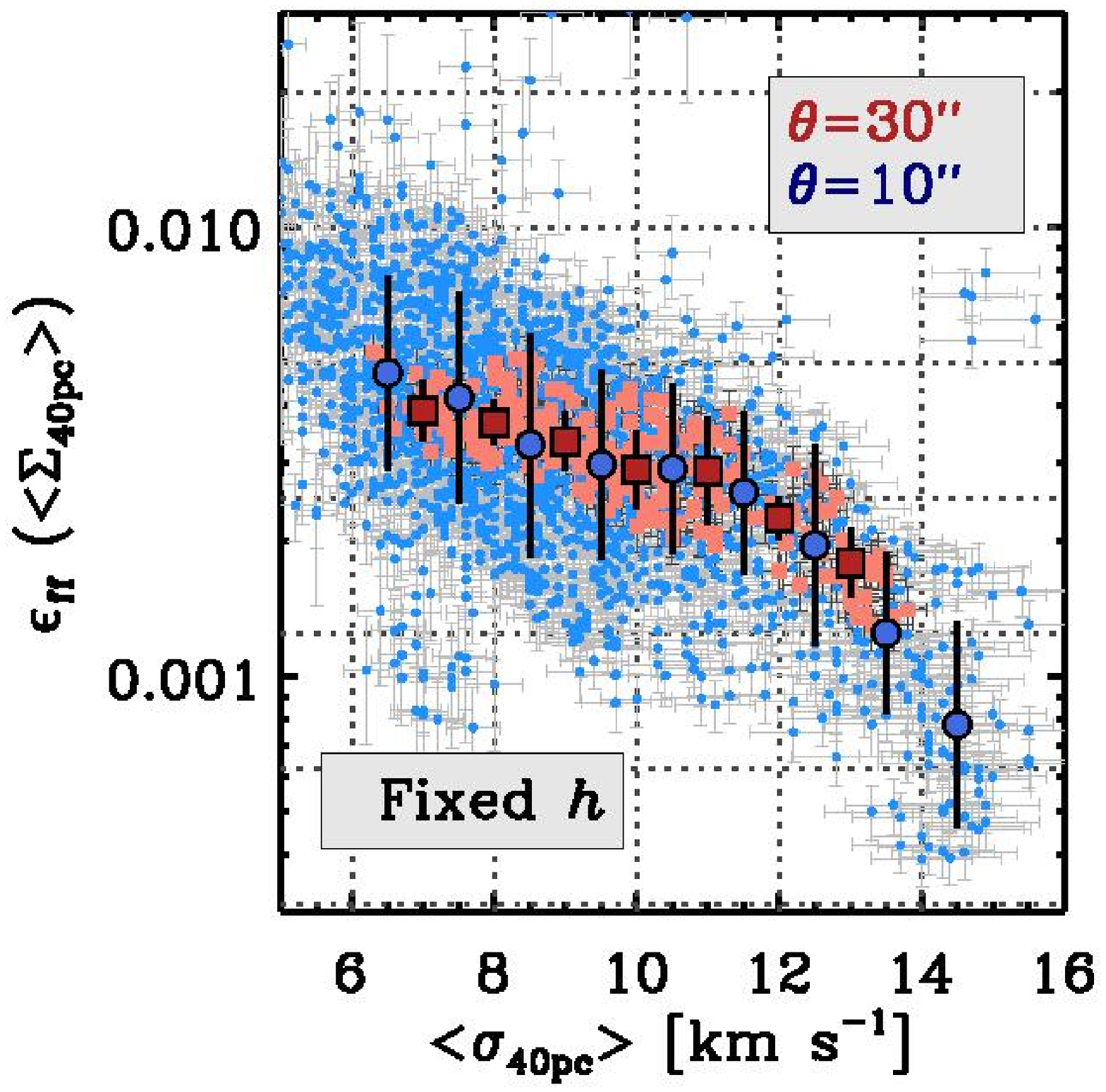}{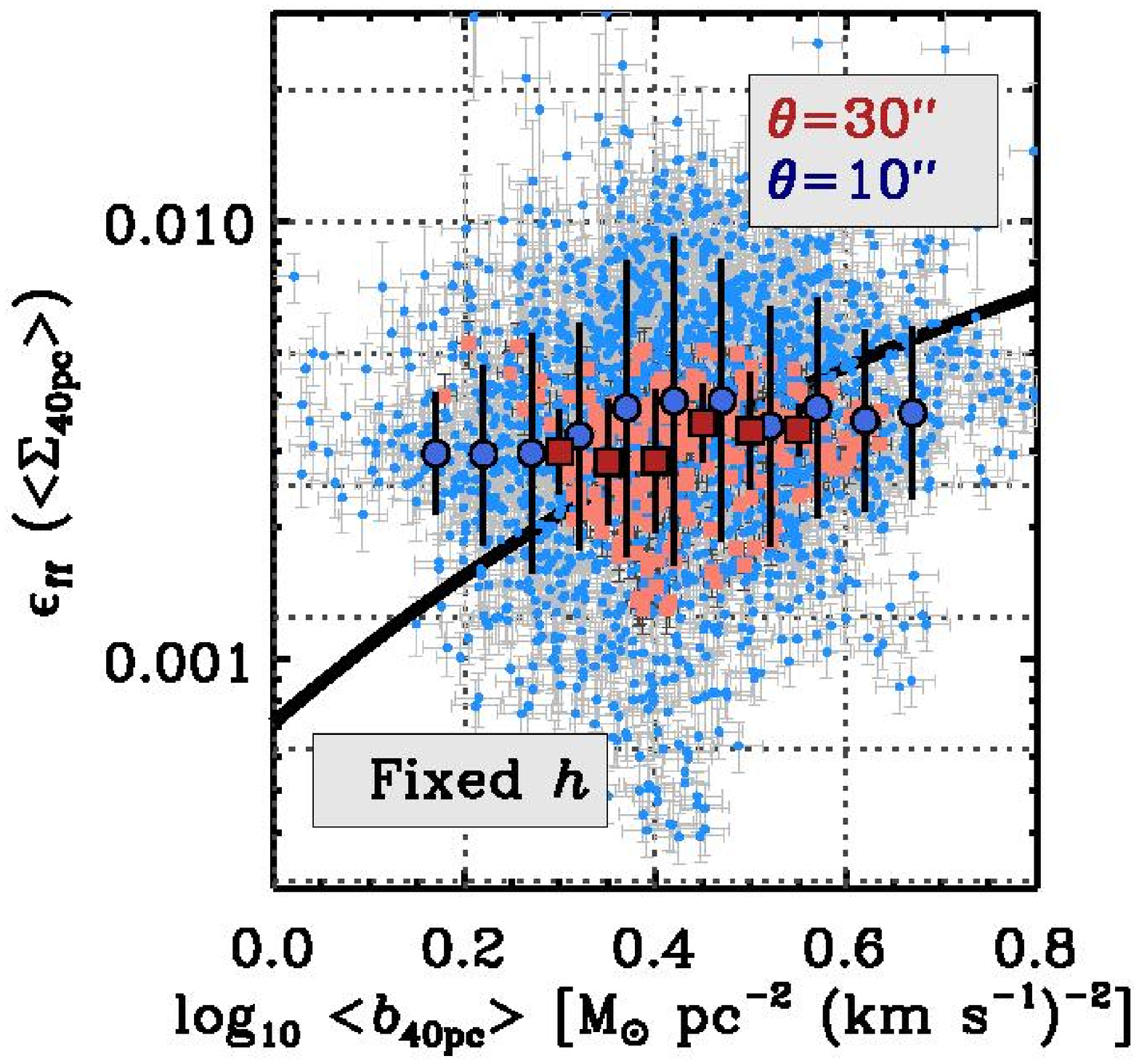}
\plottwo{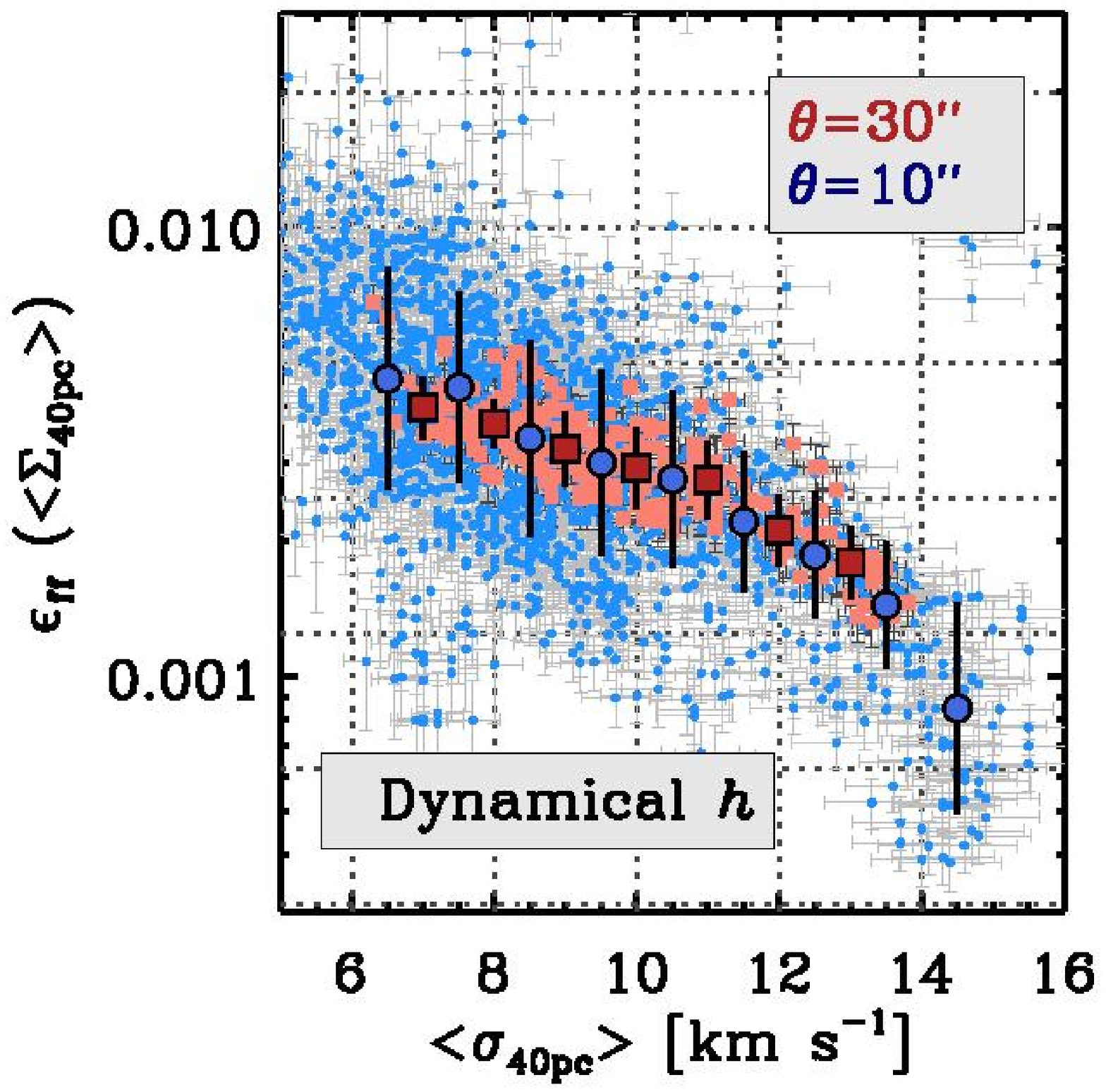}{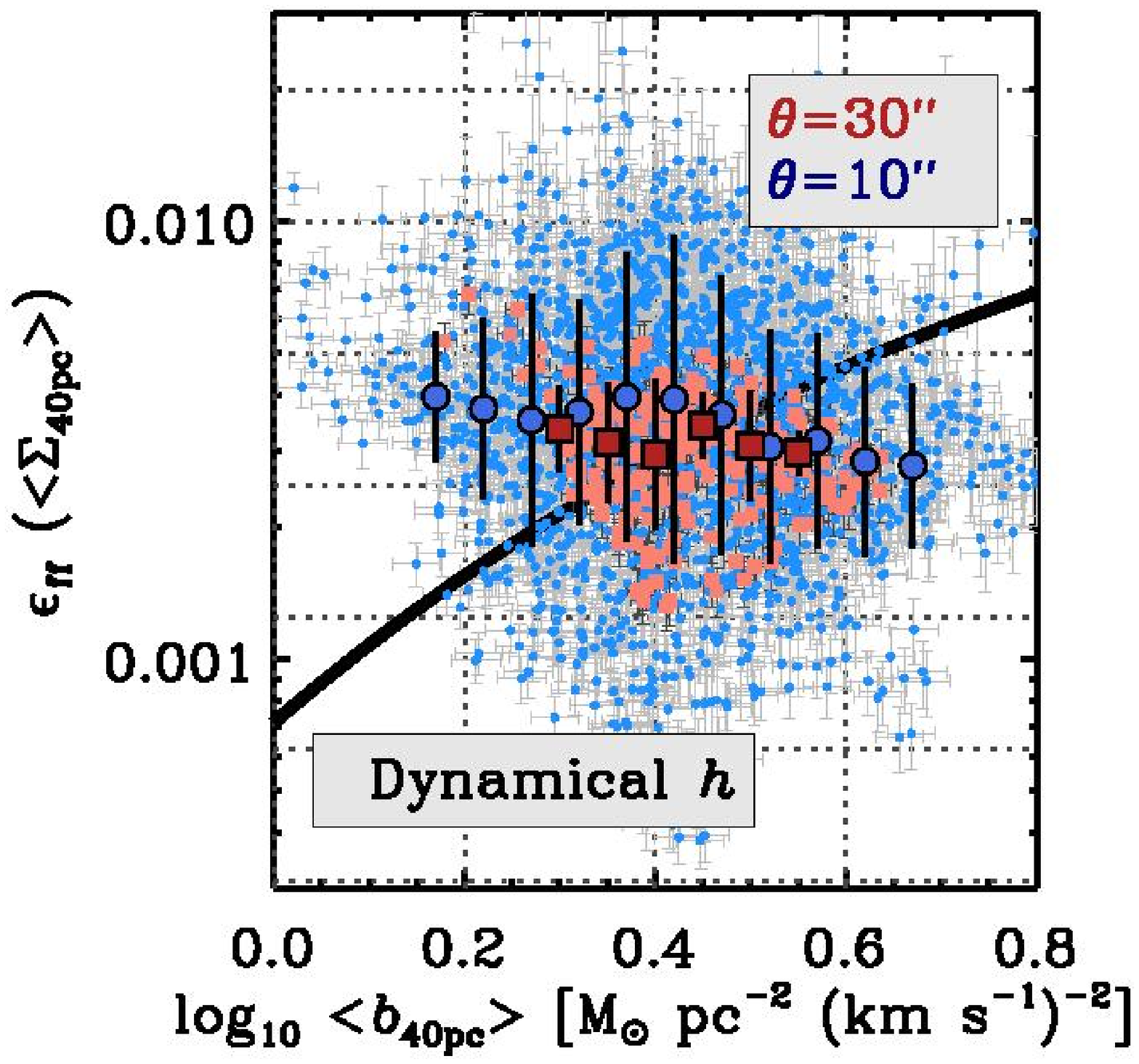}
\caption{Molecular gas depletion time, $\tau_{\rm Dep}^{\rm mol}$ (top row) and  efficiency per free-fall time, $\effsdavg$ (middle and bottom row) as a function of ({\em left})  small-scale velocity dispersion, $\sigavg$ and ({\em right}) $\bavg \equiv \Sigma/\sigma^2 \propto \alpha_{\rm vir}^{-1}$, a tracer of the dynamical state of the gas. ({\em Top left}) $\tau_{\rm Dep}^{\rm mol}$ shows little relation to $\sigavg$ at low $\lesssim 12$~km~s$^{-1}$ values. In regions with higher line widths, $>12$~km~s$^{-1}$, we find high $\tau_{\rm Dep}^{\rm mol}$, indicating a low rate of star formation per unit gas mass. ({\em Top right}) $\tau_{\rm Dep}^{\rm mol}$ anti-correlates with $\bavg$, indicating a higher rate of star formation per unit gas mass for regions with stronger self-gravity (high $b$, low $\alpha_{\rm vir}$). ({\em Middle and bottom left}) The efficiency per free-fall time anti-correlates with the line width across the galaxy, with much lower $\effsdavg$ in gas with very large line widths. The same result holds for a fixed line of sight depth $h=100$~pc or a line of sight depth that varies $h \propto b^{-1}$. ({\em Middle and bottom right}) $\effavg$ weakly correlates with $\bavg$ for the case of a fixed scale height. This becomes a weak anti-correlation if we take $h \propto b^{-1}$; that is, if we assume a fixed dynamical state and use the measured $b$ to infer $h$. The anti-correlation of $\tau_{\rm Dep}^{\rm mol}$ with $\bavg$ is much stronger than that with $\sdavg$; this offers strong, though still indirect, support to the interpretation of the top right panel as a dynamical effect, not a line of sight depth effect. The black line in the middle and bottom right panels shows $\effsdavg \propto \exp (-1.6~(5.5/b)^{0.5})$, approximately the expectation from \citet{PADOAN12}, with the normalization chosen to intersect our data.}
\label{fig:kin}
\end{figure*}

Surface density and volume density not the only relevant properties of the gas. In a turbulence-regulated view of star formation, clouds with a high Mach number has a wider density distribution and include more dense gas \citep{PADOAN02}. The Mach number also affects the critical density for the onset of star formation \citep[e.g.,][]{KRUMHOLZ05}, with a higher threshold density expected for higher Mach numbers. 

Specific predictions differ from model to model \citep[see][]{FEDERRATH12}, but most models predict an increase in $\epsilon_{\rm ff}$ for high $\mathcal{M}$. If the temperature does not vary strongly across M51, and if the line widths that we observe are primarily turbulent in nature, then \sigavg\ should reflect the turbulent Mach number. In this case, if the turbulent models are right, then we would expect \effsdavg\ to correlate with \sigavg . 

We test these expectations in the left panels of Figure~\ref{fig:kin}. We plot $\tau_{\rm Dep}^{\rm mol}$ (top) and \effsdavg\ (middle and bottom) as a function of $\sigavg$. We do not observe a significant correlation between $\sigavg$ and $\tau_{\rm Dep}^{\rm mol}$ at intermediate values of $\sigavg \approx 6{-}12$ km~s$^{-1}$. At high values of $\sigavg$, we tend to find higher $\tau_{\rm Dep}^{\rm mol}$. That is, where $\sigavg$ appears high, gas appears inefficient at forming stars.

Normalizing by the free-fall time, the middle and bottom left panels of Figure~\ref{fig:kin} show a steady decrease in $\effsdavg$ with increasing $\sigavg$. The decline becoming steeper at high $\sigavg$. The trend remains qualitatively the same for both treatments of line of sight depth. This anti-correlation is unexpected in turbulent theories. It suggests that the primary impact of the measured line width, whatever its origin, is to offer increased support against collapse rather than to increase the abundance of dense gas.

Based on modeling the velocity field, \citet{MEIDT13} and \citet{COLOMBO14B} suggested that the line widths in M51 include substantial contributions from unresolved bulk motions. In this case, \sigavg\ may instead indicate the strength shearing or streaming motions, which can play a key role suppressing star formation \citep{MEIDT13}. This seems very likely to explain the long depletion times at high \sigavg\ ($\gtrsim 12$~km~s$^{-1}$). 

At lower \sigavg\ the picture is less clear. M51 obeys the standard GMC scaling relations \citet{COLOMBO14A}, including when analyzed beam-by-beam \citet{LEROY16}, so we do expect that over most of the galaxy \sigavg\ reflects the turbulent line width to a reasonable degree (though see S. Meidt et al. submitted). In this case, Figure~\ref{fig:kin} presents a result not expected in turbulent theory: that high line width implies a low efficiency per free fall time. Making similar measurements in other galaxies will help illuminate whether this effect is general or indeed driven by the large scale dynamics of M51.

\subsubsection{Dynamical State}

Neither the surface density nor the line width exist in a vacuum. Instead, they correlate \citet[see][]{LEROY16}, so that the high $\tau_{\rm Dep}^{\rm mol}$, high \sigavg\ points in Figure~\ref{fig:kin} are also the high surface density points seen above. Their balance, $\Sigma / \sigma^2$, reflects the relative strength of the gravitational potential and the kinetic energy of the gas. In almost any view of star formation, a higher degree of self-gravity will render gas better at forming stars. In turbulent theories, this manifests as a dependence of $\epsilon_{\rm ff}$ on the virial parameter \citep[e.g.,][]{KRUMHOLZ05}, or the closely related ratio of free-fall time to crossing time \citep[e.g.,][]{PADOAN11}.

We capture the balance of gravitational potential and kinetic energy via $\bavg \equiv \sdavg / \sigavg^2 \propto {\rm UE/KE} \propto \alpha_{\rm vir}^{-1}$. When $\bavg$ is high, the surface density is high relative to the line width and the gas more tightly bound; when $\bavg$ is low it has a large kinetic energy compared to its inferred potential. 

The right panels in Figure~\ref{fig:kin} show $\tau_{\rm Dep}^{\rm mol}$ (top) and \effsdavg\ (middle and bottom) as a function of $\bavg$. We observe a significant anti-correlation between $\tau_{\rm Dep}^{\rm mol}$ and $\bavg$. The sense of this anti-correlation is that more bound gas (high $b$) --- equivalently, gas with a high ratio of $\tau_{\rm ff}$ to $\tau_{\rm cross}$ --- forms stars at a high rate per unit gas mass (low $\tau_{\rm Dep}^{\rm mol}$). The strength of the anti-correlation is striking given the weak and inconsistent relationships between $\tau_{\rm Dep}^{\rm mol}$ and $\sdavg$ or $\sigavg$. A fit to the data treating $\bavg$ as the independent variable and using the form $\tau_{\rm Dep}^{\rm mol} \propto \bavg^{-\alpha}$ gives $\alpha = -0.8$~to~$-1.0$, with the range depending moderately on the resolution and approach used to determine the best-fitting relationship.

{\em $\bavg$ Probably Does Reflect Dynamical State:} As discussed above, $b$ can be interpreted in two ways. If the line of sight depth remains constant, then $b$ traces the dynamical state of the gas, $b \propto \alpha_{\rm vir}^{-1}$. Alternatively, if the dynamical state of the gas remains fixed, e.g., if all gas is marginally bound or virialized, then $b$ indicates the line of sight depth, with $h \propto b^{-1}$. 

Figure~\ref{fig:kin} offers a strong, if indirect, argument that variations in \bavg\ do mainly reflect changes in the dynamical state. Compare the clear, steep anticorrelation between $\tau_{\rm Dep}^{\rm mol}$ and $b$ to the weak relation between $\tau_{\rm Dep}^{\rm mol}$ and $\sdavg$ seen in Figure~\ref{fig:tdep_sd}. If the density of gas is the only variable relevant to star formation, then we would expect the two figures to show similar relations because $\rho \propto \Sigma / h$. Instead, only $b$ shows a strong anti-correlation with $\tau_{\rm Dep}^{\rm mol}$. More, the slope of the anti-correlation is $\sim -0.8$~to~$-1.0$, steeper than the slope of $-0.5$ expected from only $\tau_{\rm Dep}^{\rm mol} \propto \rho^{-0.5}$.

{\em $\bavg$ and $\effsdavg$:} The importance $b \sim \alpha_{vir}^{-1}$ has been highlighted by \citet{PADOAN12} and others \citep[e.g.,][]{KRUMHOLZ05}. Gas with a lower virial parameter and a higher UE/KE or $b$ is expected to be better at forming stars. Our result broadly supports these expectations.

Turbulent theories often predict an impact of $\alpha_{\rm vir}$ on the efficiency per free-fall time, however, not the gas depletion time. The middle and bottom right panels of Figure~\ref{fig:kin} show $\effsdavg$ as a function of $\bavg$. There, the impact of $b$ is less clear. Formally, we find a weak but significant positive correlation if we hold $h$ fixed, so that \effsdavg\ is higher with higher \bavg . But the figure shows that this is a modest effect, and the trend reverses if we allow $h$ to vary. 

\citet{PADOAN12} predict $\epsilon_{\rm ff} \approx 0.5~\exp(-1.6 \tau_{\rm ff}/\tau_{\rm cross}) \propto \exp (-1.6 b^{-0.5})$. We show a modified version of this prediction as a black line in the figures. We take $\alpha_{\rm vir} = 5.5 / b$, appropriate for clouds with $R \sim 60$~pc, and set the normalization to pass through our data. Similar to the results of \citet{LEE16} in the Milky Way, the \citet{PADOAN12} prediction does not seem to capture the full set of physics at play in our data. Though we show in the next section that it offers a better match to the data for individual dynamical regions.

\subsection{Relation to Galaxy Structure}
\label{sec:region}

\begin{figure*}
\plottwo{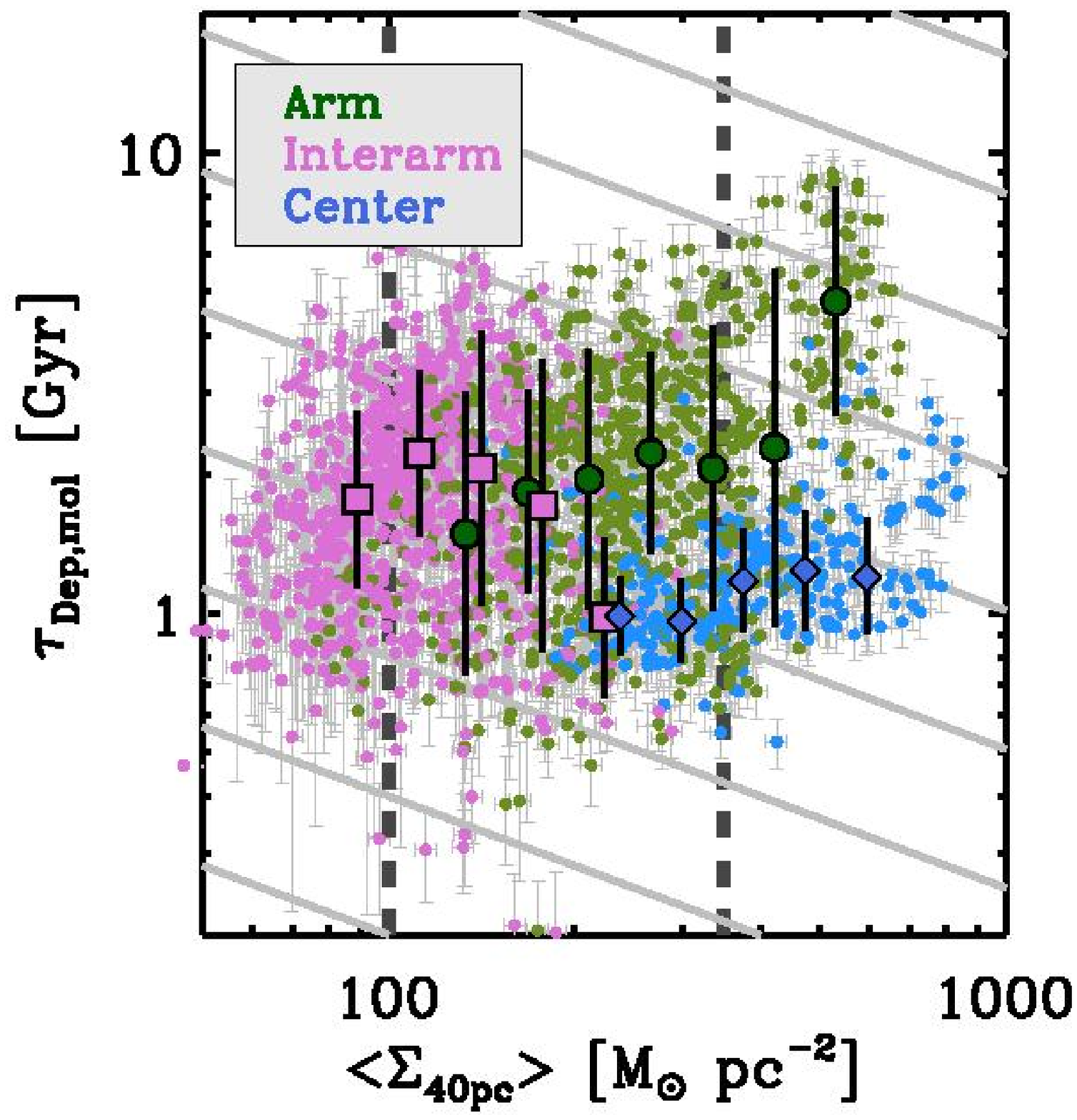}{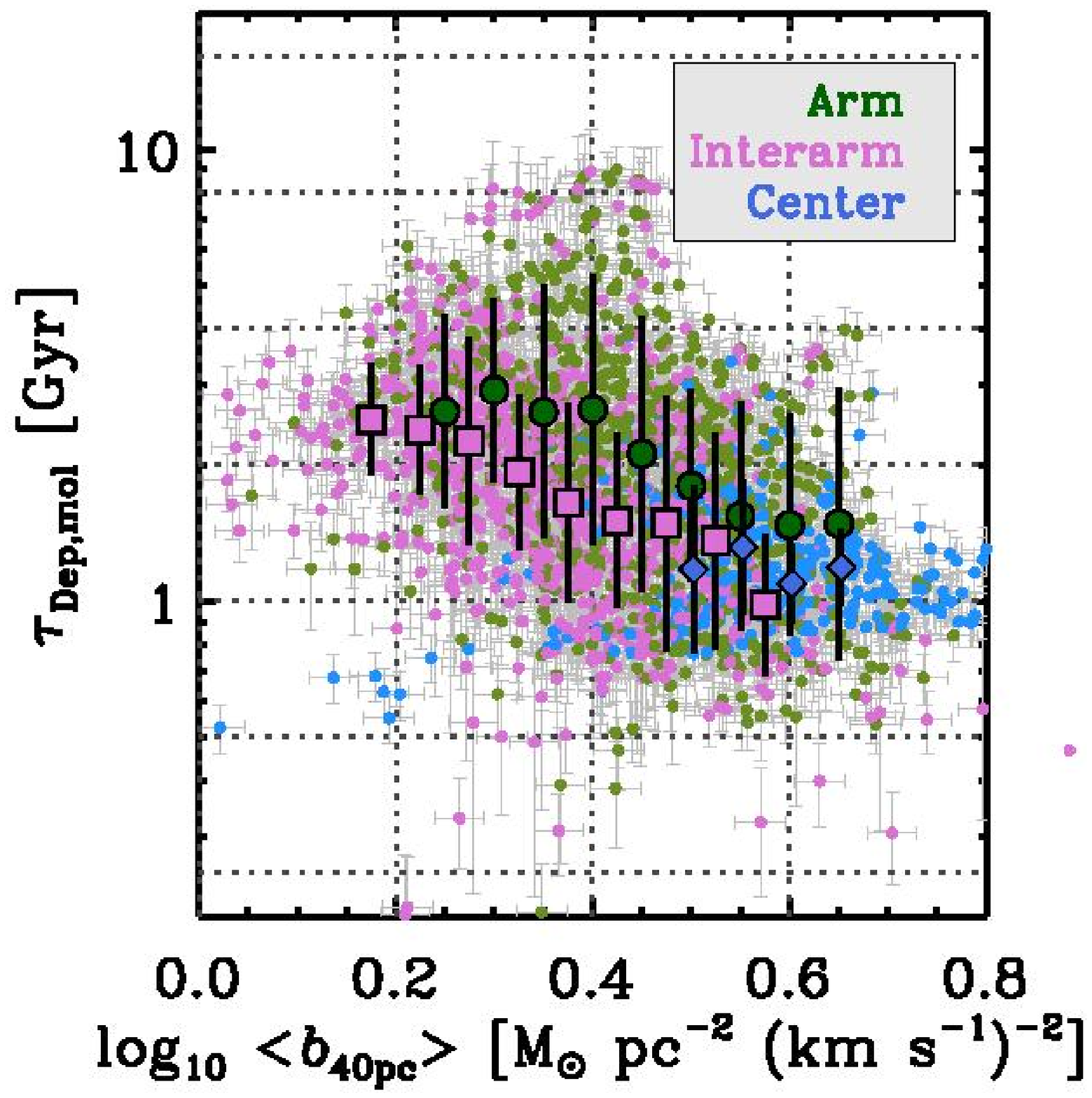}
\plottwo{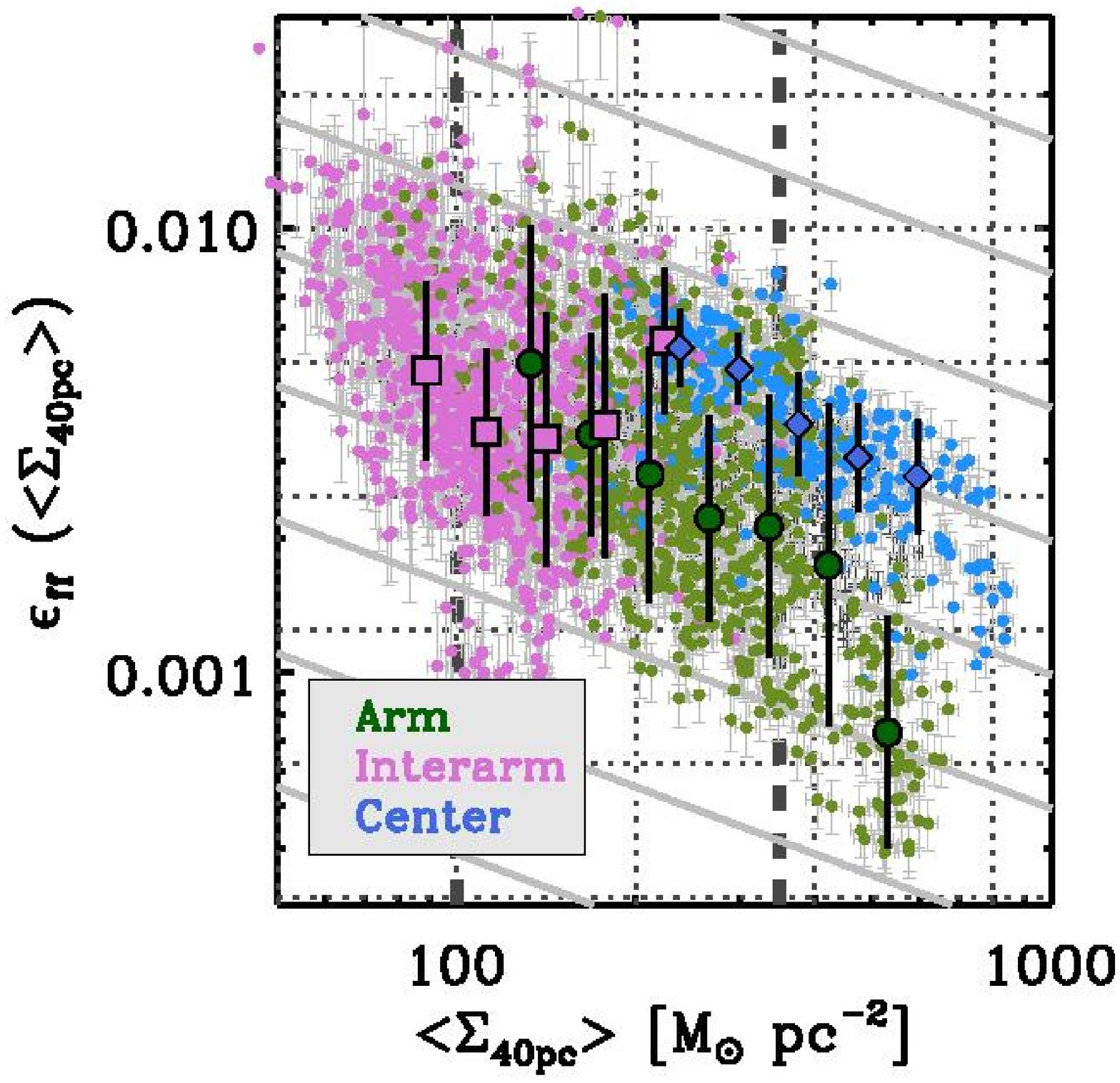}{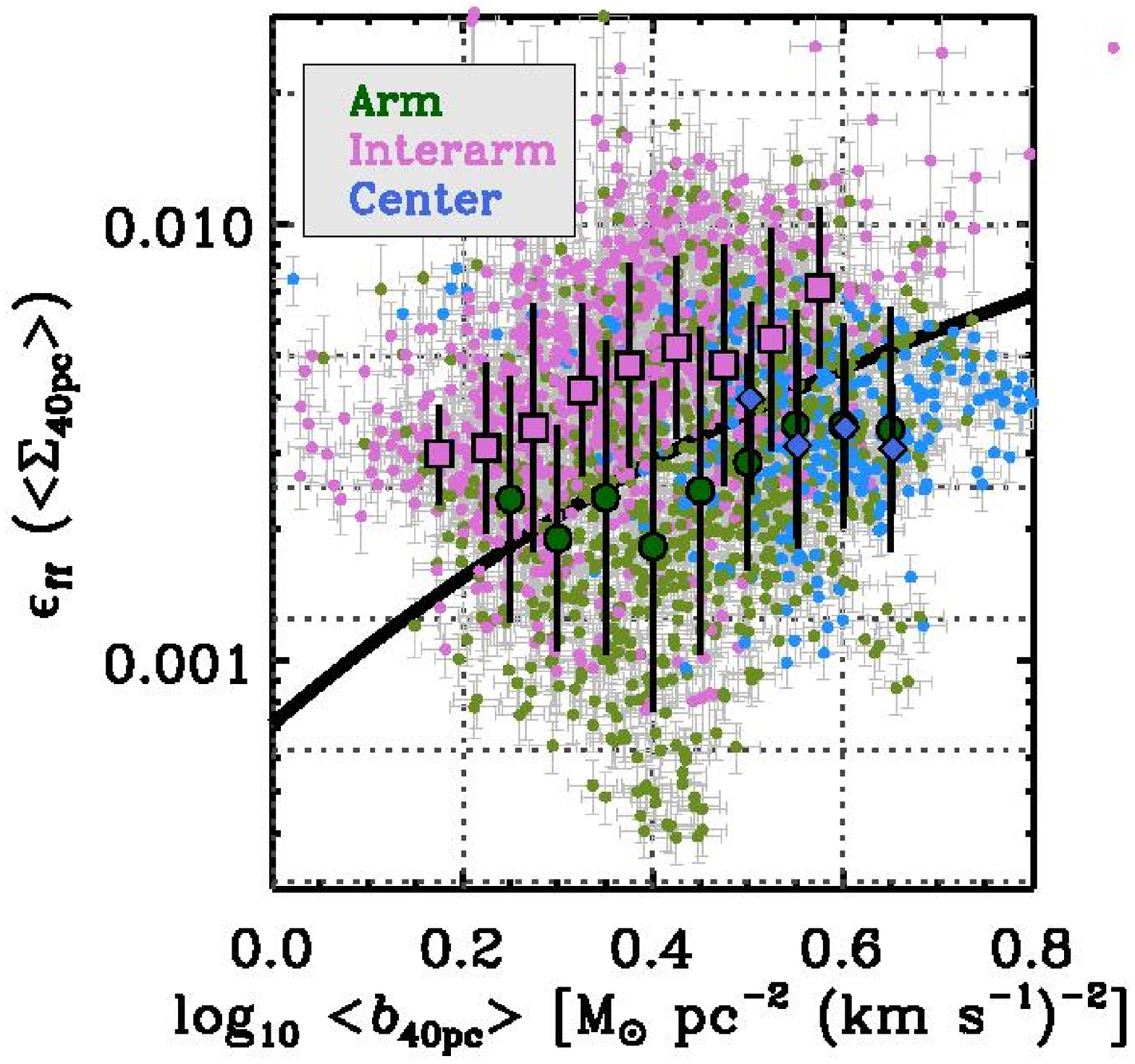}
\caption{({\em Top}:) Molecular gas depletion time, $\tau_{\rm Dep}^{\rm mol}$, and ({\em bottom}) efficiency per free fall time, $\effsdavg$ , for $h=100$~pc as a function of $\sdavg$ and $\bavg$ (as in Figures~\ref{fig:tdep_sd} and~\ref{fig:kin}) all at $10\arcsec \approx 370$~pc resolution, but now plotting measurements from the arm (green), interarm (purple), and central (blue) regions of the galaxy separately. Gray lines again show the expectation for fixed efficiency per free-fall time. The galaxy separates by region in the $\tau_{\rm Dep}^{\rm mol}$ vs. \sdavg\ diagram. The behavior of different regions appears more similar in $\tau_{\rm Dep}^{\rm mol}$ vs. $b$, consistent with the dynamical state of the gas explaining most of the observed variations in $\tau_{\rm Dep}^{\rm mol}$. Considering \effsdavg\ as a function of $\bavg$ (bottom right panel), individual regions show more indication than the galaxy as a whole for an expected positive correlation between $\effsdavg$ and $b \equiv \Sigma/\sigma^2 \propto \alpha_{\rm vir}^{-1}$. The black line in the bottom right panels shows $\effsdavg \propto \exp (-1.6~(5.5/b)^{0.5})$, approximately the expectation from \citet{PADOAN12}, with the normalization chosen to intersect our data.}
\label{fig:region}
\end{figure*}

\begin{figure*}
\plotone{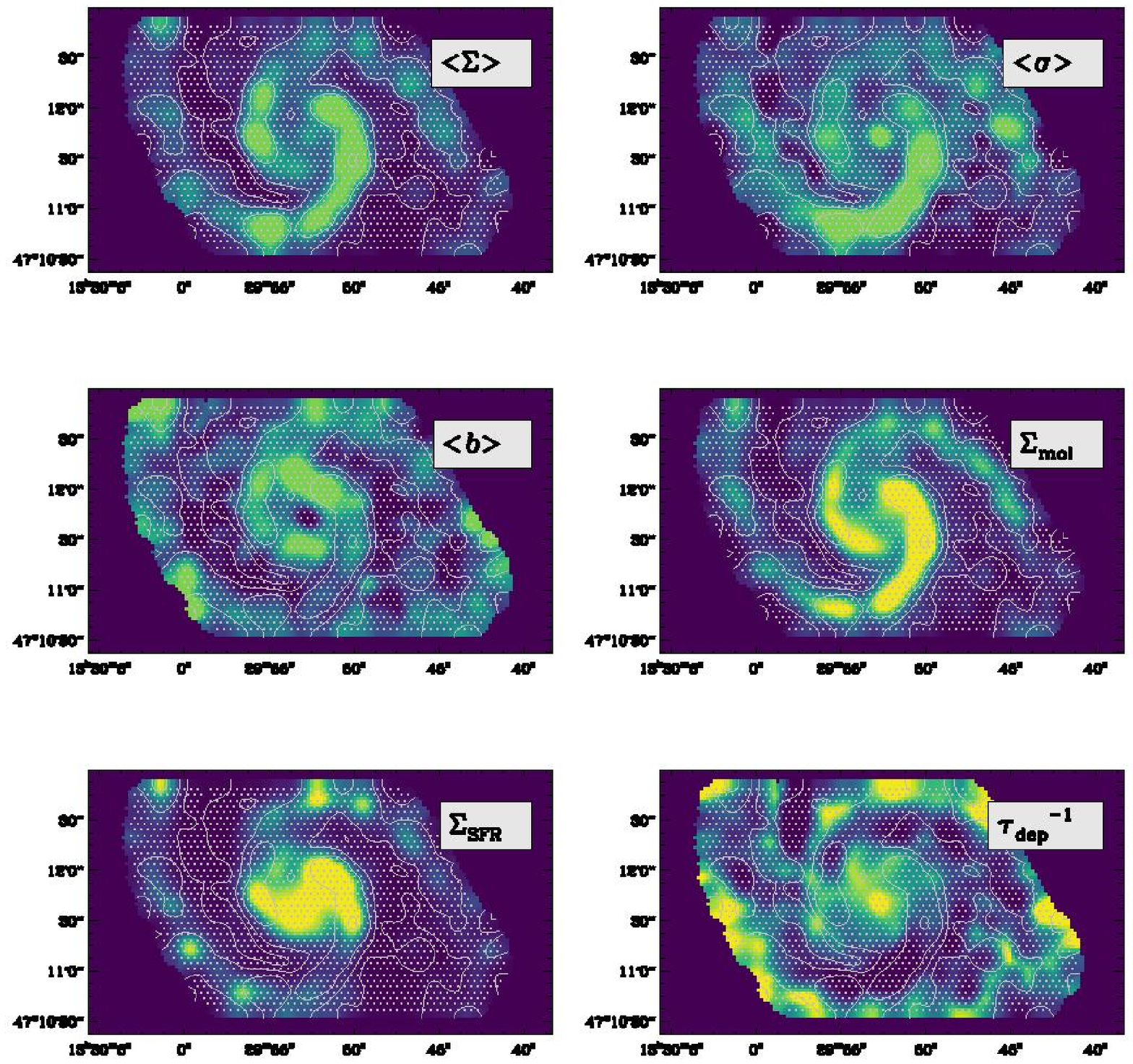}
\caption{Maps of $\sdavg$, $\sigavg$, $\bavg$, $\Sigma_{\rm mol}$, $\Sigma_{\rm TIR} \propto \Sigma_{\rm SFR}$, and $(\tau_{\rm Dep}^{\rm mol})^{-1} \equiv \Sigma_{\rm SFR}/\Sigma_{\rm mol}$ at $10\arcsec \approx 370$~pc resolution. The same contours of \sdavg\ appear in all of the images, and all images are stretched to show a linear stretch covering the middle 95\% of the data. The star-forming ring and the outer spiral arm regions show high rates of star formation per unit gas mass, and also high $b$. The inner spiral arms show high $\Sigma_{\rm mol}$ and $\sdavg$, but even higher $\sigavg$, leading to a low $b$ and comparatively weak star formation.}
\label{fig:maps}
\end{figure*}

M51 exhibits strong spiral and radial structure. Large scale gas flows have been linked to the ability of M51's gas to form stars \citep[e.g.,][]{KODA09} and to the suppression of star formation by streaming motions \citep[e.g.,][]{MEIDT13}. Figure~\ref{fig:region} shows how $\tau_{\rm Dep}^{\rm mol}$ and \effsdavg\ vary with $\sdavg$ and $\bavg$ region-by-region. Here, we color the points according to the dynamical region from which most of the CO emission in the beam originates. We show the two dimensional distributions of \sdavg, \sigavg, \bavg, $\Sigma_{\rm mol}$, $\Sigma_{\rm SFR}$, and $(\tau_{\rm Dep}^{\rm mol})^{-1}$ in Figure~\ref{fig:maps}.

Figure \ref{fig:region} shows $\tau_{\rm Dep}^{\rm mol}$ as a function of \sdavg\ for arm (green), interarm (purple), and central (blue) parts of the galaxy. As previously shown by \citet{KODA09}, \citet{HUGHES13B}, and \cite{COLOMBO14A}, the cloud-scale surface density increases dramatically moving from the interarm to arm region. The center of the galaxy exhibits high gas surface densities.

Although the arms concentrate molecular gas, we do not observe a decrease in $\tau_{\rm Dep}^{\rm mol}$ moving from the interarm to arm regions. Combining the arm and interarm regions, $\tau_{\rm Dep}^{\rm mol}$ remains approximately constant as a function of surface density until it rises at the highest values of \sdavg . This is the apparent suppression of star formation --- despite high surface densities --- observed in the arms by \citet{MEIDT13}. These observations are also consistent with the observation by \citet{FOYLE10} of a weak contrast in $\tau_{\rm Dep}^{\rm mol}$ between arm and interarm regions in M51. 

The inner part of M51 has high $\sdavg$, similar to that found in the spiral arms. Here, however, the high surface densities are accompanied by low $\tau_{\rm Dep}^{\rm mol}$.  As a result, in the top left panel of Figure~\ref{fig:region} the points at high $\sdavg$ separate in $\tau_{\rm Dep}^{\rm mol}$ according to the region from which they arise. As Figure~\ref{fig:maps} shows, many of the lowest $\tau_{\rm Dep}^{\rm mol}$ arise from the star-forming ring of the galaxy. These correspond to the high $\Sigma_{\rm SFR}$ points in the scaling relations in Figure~\ref{fig:scaling_law}. The few points at the galaxy center, in which AGN contamination \citep{QUEREJETA16} and beam smearing (e.g., S.~Meidt et al., in preparation) contribute most, has little effect on the overall trend.

The top right panel of Figure~\ref{fig:region} shows that although the parts of the galaxy separate in $\tau_{\rm Dep}^{\rm mol}$ vs. $\sdavg$ space, they overlap much better when $\tau_{\rm Dep}^{\rm mol}$ is plotted as a function of \bavg . That is, the long depletion times observed at high $\sdavg$ in the arms appear to be there because that gas has low \bavg , i.e., it appears weakly gravitationally bound. We observe an anti-correlation between $\tau_{\rm Dep}^{\rm mol}$ and $b$ in both the arm and interarm regions. The central region, which has the lower $\tau_{\rm Dep}^{\rm mol}$, also has the strongest self-gravity, traced by $b$.

We do observe an offset between the median $\tau_{\rm Dep}^{\rm mol}$ in the arm and interarm region at fixed $b$. At the same \bavg , points in the arms have typically $0.13$~dex ($\sim 35\%$) longer $\tau_{\rm Dep}^{\rm mol}$. This could reflect evolutionary effects on scales larger than our averaging beam. For example, \citet{SCHINNERER17} show the formation of stars along spurs displaced downstream from the arms. Or it could be driven additional suppression of star formation in the arms by dynamical effects not captured by $\bavg$ \citep{MEIDT13}. Alternatively, it could reflect a lower filling fraction in the interarm region, so that beam dilution affects the interarm points more, lowering $b$ relative to its true value. It could also reflect a low level bias in our SFR tracers, which affects the lower-magnitude $\Sigma_{\rm SFR}$ in the interarm more than in the arm. 

When recast from $\tau_{\rm Dep}^{\rm mol}$ to \effsdavg\ as a function of \sdavg\ (bottom row), the galaxy again separates. Here the arms appear as outliers. They show low \effsdavg , significantly lower than the interarm region or the center. That is, given the high surface densities in the arms, we would expect collapse to proceed quickly. But the observed $\tau_{\rm Dep}^{\rm mol}$ does not support this expectation. The contrast between these low \effsdavg\ in the arms and the higher values in the interarm regions drive the anti-correlation between \effsdavg\ and \sdavg\ observed across the whole galaxy.

The bottom right panel of Figure~\ref{fig:region} shows $\effsdavg$ as a function of \bavg\ region-by-region. When we considered the whole galaxy (Figure \ref{fig:kin}), only a weak correlation related \effsdavg\ to \bavg. Here the individual regions show a stronger positive correlation between \effsdavg\ and \bavg . There is some indication that at least the interarm regions match the sense of the \citet{PADOAN12} prediction (the black line). The picture for the arm and center regions is less clear. Together they may show a weak positive correlation between \effsdavg\ and \bavg , but it is not clear that they should be grouped together. The offset between the interarm and arm regions at fixed \bavg\ appears even stronger in \effsdavg\ than for $\tau_{\rm Dep}^{\rm mol}$. At fixed \bavg\ interarm regions have typically $\sim 0.24$~dex, almost a factor of two, higher \effsdavg\ than arm regions with the same \bavg .

Together, Figures~\ref{fig:region} and~\ref{fig:maps} paint a picture of M51 that qualitatively resembles that seen in many barred galaxies: despite the high surface densities in the inner dynamical features (here the arms), gas in this region appears stabilized against collapse. But flows along the arms feed gas condensations (the star-forming ring) in the inner regions \citep[see][]{QUEREJETA16}, where star formation activity does proceed at a high level in both an absolute and normalized sense. Despite our averaging over moderately large ($370$~pc) areas, timescale effects may also be at play. The $\tau_{\rm Dep}^{\rm mol}$ map in Figure~\ref{fig:maps} shows significant azimuthal structure, and as shown by \citet{SCHINNERER17}, star formation tends to occur in spur-like structures downstream of the arms. We refer the reader to extensive discussions in \citet{MEIDT13}, \citet{COLOMBO14A}, and \citet{QUEREJETA16}, \citet{SCHINNERER17}, and references therein, for more discussion.

\section{Discussion and Summary}
\label{sec:discuss}

We have used the PAWS survey \citep{SCHINNERER13} to compare cloud-scale ISM structure to the locally-averaged ability of gas to form stars across the inner part of M51. We compare infrared emission, tracing molecular gas mass, to recent infrared emission, tracing the recent SFR, within each $10\arcsec \approx 370$~pc and $30\arcsec \approx 1.1$~kpc beam. Then, we use the method described by \citet{LEROY16} to calculate the mass-weighted 40-pc surface density ($\sdavg$), line width ($\sigavg$), and self-gravity ($\bavg$, $b \equiv \Sigma/\sigma^2 \propto \alpha_{\rm vir}^{-1}$) in each larger beam. This is similar to recording the mass-weighted mean GMC properties in each beam, but these intensity-based measurements are simpler and require fewer assumptions than estimating cloud properties. Still they capture the key physics in the Larson scaling relations well \citep{LARSON81}.

We adopt simple translations between observed and physical quantities, so that our key results can be easily phrased in either observable or physical terms. Comparing CO and IR at large scales, we find:

\begin{enumerate}
\item At large scales, our CO and IR measurements qualitatively match previous studies of SFR-gas scaling relations in M51. The sublinear behavior noted by \citet{SHETTY13} at large radii, the superlinear behavior noted by \citet{LIU11} in the inner galaxy, and the wide range of depletion times at high gas surface density found by \citet{MEIDT13} are all evident in Figure~\ref{fig:scaling_law}.
\end{enumerate}

\noindent At $370$~pc resolution, we still observe appreciable ($\sim 0.3$~dex) scatter in the CO-to-IR ratio, rising $\sim 0.4$~dex at the highest surface densities. We compare the measured CO-to-IR ratio expressed as a molecular gas depletion time, $\tau_{\rm Dep}^{\rm mol}$, to the small scale gas structure measured from PAWS to investigate if and how local gas structure drives depletion time variations. The most basic expectation, e.g., following \citet{KRUMHOLZ12B}, is that variations in $\tau_{\rm Dep}^{\rm mol}$ result from variations in the cloud-scale density, which sets the local gravitational free fall time, $\tau_{\rm ff}$. To test this, we compare $\tau_{\rm Dep}^{\rm mol}$ to \sdavg , the mean cloud scale surface density in the beam and our best observational tracer of the gas density. We find that

\begin{enumerate}
\setcounter{enumi}{1}
\item The CO-to-IR ratio, tracing $\tau_{\rm Dep}^{\rm mol}$, shows a weak anti-correlation with \sdavg\ over the range $\sdavg \approx 100{-}350$ M$_\odot$~pc$^{-2}$ (Figure~\ref{fig:tdep_sd}). Over this range, denser gas does appear moderately better at forming stars. The slope of this anti-correlation, $\sim -0.25$~to~$-0.35$, is shallower than what is naively expected for a fixed efficiency per free-fall time.
\end{enumerate}

\noindent With an estimate of the line of sight depth, $h$, our $\sdavg$ can be translated to a density, $\rhosdavg$, and then to a gravitational free fall time, $\tau_{\rm ff}$. Contrasting $\tau_{\rm Dep}^{\rm mol}$ and $\tau_{\rm ff}$ yields an estimate of the efficiency of star formation per free fall time, a central quantity for many recent theories of star formation. We consider what line of sight depth to use based on both recent GMC catalogs and studies of the disk thickness in M51 and the Milky Way.

\begin{enumerate}
\setcounter{enumi}{2}
\item In recent GMC catalogs targeting the Milky Way \citep{HEYER09,MIVILLE17} and M51 \citep{COLOMBO14A} the volume density and surface density of clouds correlate well (Figure \ref{fig:clouds}). In these catalogs, most of the CO emission arises from clouds with $R \sim 30{-}100$~pc. The observable cloud scale surface density does appear to be a reasonable proxy for the local mean volume density, though more work is needed on this topic.
\end{enumerate}

\noindent We adopt both a fiducial depth $h=100$~pc (our best estimate) and a ``dynamical'' depth calculated from holding the virial parameter constant. For both cases, we calculate the distribution of efficiency per free fall time, \effsdavg , across the PAWS field.

\begin{enumerate}
\setcounter{enumi}{3}

\item At both of our working resolutions, $\effavg$ estimated in this way is ${\sim} 0.3{-}0.36\%$, with ${\sim} 0.3$~dex scatter for a $370$~pc averaging beam, and ${\sim} 0.1$~dex scatter for a $1.1$~kpc averaging beam (Figure~\ref{fig:eff}).

\end{enumerate}

\noindent This value agrees in broad terms with what one would infer based on comparing average GMC properties in the Milky Way and nearby galaxies\citep[e.g.,][]{BOLATTO08,HEYER09} to large-scale measurements of the molecular gas depletion time \citep[e.g.,][]{LEROY13}. It also matches the apparent requirements for turbulent models to match observations of dense gas, IR, and CO in nearby galaxies \citep{GARCIABURILLO12,USERO15}. However, our inferred $\epsilon_{\rm ff}$ is much lower than values measured for the nearest molecular clouds by \citet{EVANS14}, \citet{MURRAY11}, or \citet{LEE16} \citep[see also][]{LADA10,LADA12}, as well as for molecular clouds orbiting the Galactic Center by \citet{BARNES17}. It is also much lower than the values commonly adopted in analytic theories and numerical simulations \citep[e.g., see][among many others]{KRUMHOLZ12B,AGERTZ15}.

The main drivers for the mismatch with \citet{LEE16} and \citet{MURRAY11} appear to be sampling effects. Our method averages over all evolutionary states to calculate a regional mean $\tau_{\rm Dep}^{\rm mol}$, while their work focuses on GMCs associated with peaks of recent star formation. The discrepancy with local clouds appears more subtle, but may be an issue of matching scales; the \citet{EVANS14} measurements focus on the $A_V > 2$~mag material in local clouds, perhaps leading to the lower $\tau_{\rm Dep}^{\rm mol}$ and shorter $\tau_{\rm ff}$ in these clouds than are found at larger scales. The best ways to address these discrepancies appear to be high resolution extinction-robust estimates of the SFR, to allow experiments exactly matched to those of \citet{MURRAY11} and \citet{LEE16}, and high resolution ($\sim$ few pc resolution) CO imaging of a large area ($\sim$ kpc) in a nearby galaxy, to investigate the superstructure around analogs to the \citet{EVANS14} clouds. 

Beyond only the value of \effsdavg , we investigate how $\tau_{\rm Dep}^{\rm mol}$ and \effsdavg\ depends on the local cloud population and location in the galaxy. For $\tau_{\rm Dep}^{\rm mol}$, we find:

\begin{enumerate}
\setcounter{enumi}{4}

\item At high $\sdavg > 350$~M$_\odot$~pc$^{-2}$, the $\tau_{\rm Dep}^{\rm mol}$ increases with increasing \sdavg . This leads to the unexpected result, pointed out by \citet{MEIDT13}, that some of the highest surface density regions of M51 show relatively weak star formation. These regions lie in the spiral arms and also have high \sigavg . Their low $\tau_{\rm Dep}^{\rm mol}$ is explained, in our analysis, by the fact that this gas appears more weakly self-gravitating (lower \bavg) than other material in M51 (Figures~\ref{fig:kin} and~\ref{fig:region}).

\item Instead of either surface density or line width alone, $\tau_{\rm Dep}^{\rm mol}$ appears most closely related to the ratio $b \equiv \Sigma / \sigma^2$ (Figure~\ref{fig:kin}). Within a length scale (the line-of-sight depth through the disk), \bavg\ traces the strength of self-gravity, $b \propto \alpha_{\rm vir}^{-1} \propto {\rm UE/KE} \propto \tau_{\rm ff}^2/\tau_{\rm cross}^2$. Thus, gas that appears more gravitationally bound also appears better at forming stars. The power law slope relating $\tau_{\rm Dep}^{\rm mol}$ to $\bavg$ is $ \tau_{\rm Dep}^{\rm mol} \propto b^{\beta}$ with $\beta = {-}0.8$~to~${-}1.0$. 

\item All three regions of the galaxy (arm, interarm, and center) line up in $\tau_{\rm Dep}^{\rm mol}$-\bavg\ space, with only a modest, $0.13$~dex ($\sim 35\%$) offset (Figure \ref{fig:region}). This offset has the sense that the arm region has a modestly higher $\tau_{\rm Dep}^{\rm mol}$ (CO-to-IR) than the other regions at fixed $\bavg$.
\end{enumerate}

\noindent Our difference in results comparing $\tau_{\rm Dep}^{\rm mol}$ to \bavg\ and \sdavg\ suggest that $b$ does indeed trace dynamical state.  If both traced density, and if density represented the only important variable, then we would $\tau_{\rm Dep}^{\rm mol}$ to depend on \bavg\ and \sdavg\ in the same way. Instead, $\tau_{\rm Dep}^{\rm mol}$ show a steeper, more significant relation to $\bavg$ than to $\sdavg$.

This apparent dependence of star formation on the dynamical state of the gas, or equivalently the virial parameter, echos findings for the Milky Way. There, the largest reservoir of high-density gas in the Galaxy is also currently the least efficient at forming stars. This phenomenon is thought to be caused by shear and the supervirial nature of the clouds \citep{KRUIJSSEN14B}. Similarly, though our formalisms differ, our findings qualitatively agree with \citet{MEIDT13}, who argued that the dynamical state of the gas in M51's arms, as observed by PAWS, suppresses star formation. Our results also agrees with theoretical expectations in broad brush \citep[e.g.,][]{PADOAN12,KRUMHOLZ05,FEDERRATH12}. In detail, however, those models often make predictions about the efficiency of star formation per gravitational free fall time. We compare \effsdavg\ to the local cloud populations and find:

\begin{enumerate}
\setcounter{enumi}{8}

\item In general \effsdavg\ appears anti-correlated with \sdavg\ and \sigavg\ (Figure \ref{fig:kin}). The anti-correlation with \sdavg\ is weak over the range $\sdavg \approx 100{-}350$ M$_\odot$~pc$^{-2}$, but becomes stronger at high $\sdavg$. The anti-correlation between $\effsdavg$ and $\sigavg$ appears strong across the full range of $\sigavg$ and becomes stronger at high dispersions. In general, a higher surface density and a higher line width both appear to imply lower efficiency per free-fall time in M51 (Figures \ref{fig:eff_sd} and \ref{fig:kin}).

\item We find a weak positive correlation between \effsdavg\ and $\bavg$ for fixed line-of-sight depth, so that gas with higher apparent self-gravity appears to have a higher efficiency per free-fall time. Considering the whole galaxy, the strength of this correlation is weaker than the dependence predicted by the turbulent star formation law of \citet{PADOAN12} (Figure \ref{fig:kin}).

\item \effsdavg\ appears to correlate better with $b \propto \alpha_{\rm vir}^{-1}$ within an individual dynamical region, particularly within the interarm region (Figure \ref{fig:region}). At fixed $\bavg$, we find \effsdavg\ to be $\sim 0.24$~dex lower in the arm regions than the interarm regions, on average. Thus relative to the expected collapse time, star formation is suppressed in the arms relative to the interarms by almost a factor of two at fixed virial parameter (Figure \ref{fig:region}).
\end{enumerate}

\noindent Turbulent star formation models tend to predict a positive correlation between $\epsilon_{\rm ff}$ and the Mach number, related to our observed line width. They also tend to predict a strong dependence of $\epsilon_{\rm ff}$ on $b \propto \alpha_{\rm vir}^{-1}$. Several theories have invoked an approximately fixed \effsdavg . Thus, in detail our observations do not show outstanding agreement with current models. However, those models include a number of additional dependencies, including on factors such as the magnetic field, character of the turbulence \citep[see summary in][]{FEDERRATH12}. Our measurements also represent population, and so time, averages by design. So any dynamical cloud lifetime \citep[][]{MURRAY11,LEE16}.

To facilitate comparison with such models, we include all of our measurements in Table \ref{tab:data}. We emphasize that our intensity-based approach is easy to replicate with no need for cloud-finding or other complex image processing. Indeed, numerical simulations can directly match our line-of-sight approach and so marginalize over some of the geometrical uncertainties. Our approach to physical parameter estimation is simple and straightforward to treat via forward modeling. A main goal of this paper is to provide these measurements as an extragalactic benchmark for theories of star formation that consider cloud-scale gas structure.

Finally, as discussed in the text and appendix, there are systematic uncertainties regarding the CO-to-H$_2$ conversion factor, star formation rate, and line of sight geometry. We motivate our choices in the text and appendix and test the impact of our assumptions, but these issues are standard in this field and should be born in mind when considering the results of the paper. We also anticipate refining technical details of our weighting averaging methods over the next year to better treating ensembles of line profiles and de-emphasize the impact of an extended averaging beam \citep[see][]{LEROY16}.

\subsection{Next Steps}

Within the next year, it should be possible to conduct a similar analysis as we present here for M51 for a diverse sample of local galaxies. These include the other five galaxies treated by \citet{LEROY16} and targets of new ALMA mapping surveys that achieve cloud-scale resolution across ${\sim} 10$ star-forming galaxies. Such tests will establish: 1) if our observed very low \effsdavg\ is universal, 2) if the apparent role of self-gravity traced by $b$ is unique to M51 or a general feature, and 3) whether the gravitational free-fall time estimated from high-resolution imaging indeed appears to be a controlling parameter. Combination of these cloud-scale measurements with density-sensitive spectroscopy \citep[e.g.,][]{USERO15,BIGIEL16,LEROY17} will also help connect structural analysis at the GMC-scale to the internal density structure of clouds, which plays a key role in their ability to form stars.

Our structural analysis follows the ``beamwise'' approach described in \citet{LEROY16}, but a large literature exists estimating GMC properties for nearby galaxies \citep[e.g.,][the latter for M51]{FUKUI10,COLOMBO14A}. Following similar studies in the Milky Way \citep[e.g.,][]{MURRAY11,EVANS14}, these measurements can be compared to $\tau_{\rm Dep}^{\rm mol}$ in a similar way to what we do here. A.~Schruba et al. (in preparation) present such an analysis for a large collection of galaxies with GMC property measurements.

Finally, two major observational steps could address the tension between our measurements and those of the Milky Way. First, by observing CO from a large part of a star-forming galaxy at very high spatial resolution, one could attempt to mimic the Milky Way observations with full knowledge of the surrounding medium. Second, pairing extinction robust star formation rate tracers with high resolution gas mapping would allow the kind of population studies carried out by \citet{LEE16}. The need to leverage low resolution IR maps to estimate the star formation rate limits current efforts to consider population averages at few hundred pc scales.

\acknowledgments We thank the anonymous referee for a thoughtful and constructive report. This work is based on observations carried out with the IRAM NOEMA Interferometer and the IRAM \mbox{30-m} telescope. IRAM is supported by INSU/CNRS (France), MPG (Germany) and IGN (Spain). This work was carried out as part of the PHANGS collaboration (formerly SFNG) and the PAWS collaboration. The work of AKL, MG, and JS is partially supported by the National Science Foundation under Grants No. 1615105, 1615109, and 1653300. ES acknowledge financial support to the DAGAL network from the People Programme (Marie Curie Actions) of the European Union's Seventh Framework Programme FP7/2007- 2013/ under REA grant agreement number PITN-GA-2011-289313. ES acknowledges funding from the European Research Council (ERC) under the European Union’s Horizon 2020 research and innovation programme (grant agreement No. 694343). AH acknowledges support from the Centre National d'Etudes Spatiales (CNES). JMDK and MC gratefully acknowledge financial support in the form of an Emmy Noether Research Group from the Deutsche Forschungsgemeinschaft (DFG), grant number KR4801/1-1 (PI: Kruijssen). JMDK acknowledges funding from the European Research Council (ERC) under the European Union's Horizon 2020 research and innovation programme via the ERC Starting Grant MUSTANG (grant agreement number 714907, PI Kruijssen). GB is supported by CONICYT/FONDECYT, Programa de Iniciacion, Folio 11150220. AU acknowledges support from Spanish MINECO grants AYA2012-32295 and FIS2012-32096. FB ac- knowledges support from DFG grant BI1546/1-1. SGB thanks economic support from grants ESP2015-68964-P and AYA2016-76682-C3-2-P.

\bibliography{akl}

\begin{appendix}

Our results depend on estimates of the recent star formation rate and molecular gas mass. We adopt simple approaches to each, utilizing the total infrared (TIR) emission as a tracer of the star formation rate and adopting a Galactic $\alpha_{\rm CO} = 4.35$~\acounits\ to convert CO~(1-0) intensity in molecular gas mass surface density. \\

\section{Other Approaches to the Star Formation Rate}

\begin{figure*}
\plotone{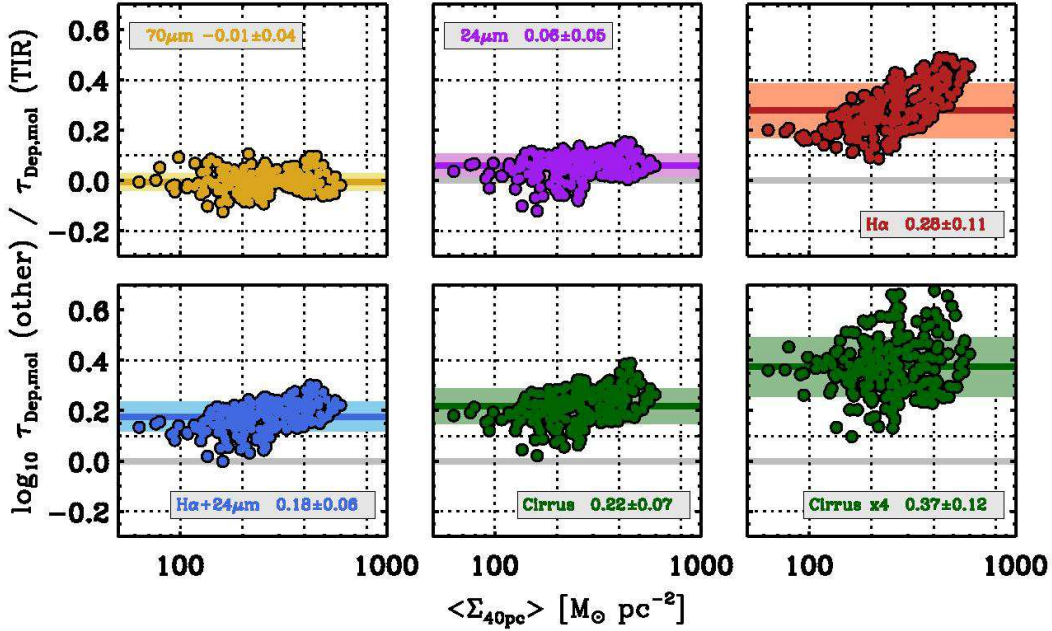}
\caption{The effect of different SFR tracers on molecular gas depletion time. At $\theta = 30\arcsec$ resolution, we estimate $\Sigma_{\rm SFR}$ using: ({\em top left}) 70 $\mu$m emission only, ({\em top middle}) 24 $\mu$m emission only, ({\em top right}) H$\alpha$ emission with one magnitude of extinction, and ({\em bottom row}) H$\alpha$+24$\mu$m with ({\em left}) no cirrus treatment, and ({\em middle} and {\em right}) $1$, and $4$ times a gas-based cirrus estimate removed from the 24 $\mu$m emission. Prescriptions follow \citet{LEROY12}, and are similar to those from \citet{MURPHY11} and \citet{CALZETTI07}. Each panel shows the ratio of $\tau_{\rm Dep}^{mol}$ measured using the other SFR tracer to what we measure based on TIR emission at $30\arcsec$ resolution. The solid line and the shaded region show the median ratio and $\pm 1\sigma$ range. Tracers involving 24 $\mu$m tend to agree well with our estimates. Using H$\alpha$ alone misses a substantial amount of extinction in regions of high gas surface density. Overall, most other estimates tend to modestly increase $\tau_{\rm Dep}^{\rm mol}$, which would imply lower $\effsdavg$. None of the alternative SFR tracers appear to induce a downward tilt in the diagram, which we would expect for a fixed $\effsdavg$.}
\label{fig:bysfr}
\end{figure*}

We use TIR intensity as our tracer of SFR. At $\theta = 30\arcsec$, we calculate $\Sigma_{\rm TIR}$ using four bands and the SED-fitting based prescription of \citet{GALAMETZ13}. At $\theta=10\arcsec$, we use a linear translation of $I_{70}$ into $\Sigma_{\rm TIR}$, with the coefficient derived from comparing 70 $\mu$m intensity to TIR intensity at $30\arcsec$ resolution. We then translate $\Sigma_{\rm TIR}$ to $\Sigma_{\rm SFR}$ following \citet{MURPHY11}.

The main impact of $\Sigma_{\rm SFR}$ in this paper is on the estimate of $\tau_{\rm Dep}^{\rm mol}$. To assess the impact of our choice of estimator, Figure \ref{fig:bysfr} shows the effect on $\tau_{\rm Dep}^{\rm mol}$ of replacing our adopted TIR-based SFR with estimates using a different approach. We only have access to all of the required data at $\theta=30\arcsec$, so this plot shows only results for that resolution over the PAWS field, our area of interest.

First, we show results using only 70 $\mu$m emission and the formulae quoted in Sections \ref{sec:meas} and \ref{sec:convtophys}. This is our approach at $\theta=10\arcsec$, where {\em Herschel}'s 70 $\mu$m map is our only available IR band. We also show results using only 24 $\mu$m, using H$\alpha$ assuming one magnitude of extinction, hybridizing H$\alpha$ and 24 $\mu$m emission, and combining H$\alpha$ with 24 $\mu$m after subtracting a ``cirrus'' (non star-forming) component from the 24 $\mu$m emission. Except for the 70 $\mu$m emission, the prescriptions used for the other tracers are taken from \citet{LEROY12}, which builds heavily on \citet{CALZETTI07} and \citet{MURPHY11}. We use the gas based cirrus prediction, which assumes a typical dust-to-gas ratio and that all of the gas is illuminated by a radiation field $0.6$ times that found in the Solar Neighborhood. The final panel shows the result for quadrupling the radiation field used in the cirrus estimate. Including an FUV-based hybrid \citep[as in][]{LEROY08} would not add much to the analysis given the heavily extinguished nature of the region in question \citep[see][]{LEROY12}. Each panel quotes the median and scatter in the logarithm of the ratio between $\tau_{\rm Dep}^{\rm mol}$ estimated using this other tracer to that used in the main body of the paper.

The figure shows that the IR-based estimates agree well with one another and yield higher $\Sigma_{\rm SFR}$ than estimates using H$\alpha$. Indeed, the main result of changing the SFR tracer is usually to lower $\Sigma_{\rm SFR}$, thereby increasing $\tau_{\rm Dep}^{\rm mol}$. The magnitude of the shift is a factor of ${\sim} 2$ if only H$\alpha$ with $1$~mag of extinction is used or a very large cirrus component is adopted (which also amounts to only weakly correcting H$\alpha$ for extinction). A main result of our analysis is a low \effsdavg . Lower $\Sigma_{\rm SFR}$ and higher $\tau_{\rm Dep}^{\rm mol}$ would drive \effsdavg\ to even lower values. In detail, given the gas-rich, dusty nature of the inner few kpc of M51, we do not necessarily expect these lower $\Sigma_{\rm SFR}$ estimates to be more correct, but if they are then it would not change our qualitative conclusions.

Note that data at higher \sdavg\ tend to show a larger discrepancy between IR-based SFR estimates and H$\alpha$ with little or no correction. The sense of this trend is that most alternatives to the IR-based $\Sigma_{\rm SFR}$ would yield longer $\tau_{\rm Dep}^{\rm mol}$ at higher \sdavg. The result would be an even lower $\effsdavg$ at high $\sdavg$ than we already observe. That is, none of the alternatives in Figure \ref{fig:bysfr} push the data towards a more nearly fixed \effsdavg . 

More, recall that Figure \ref{fig:scaling_law} shows that our IR based approach yields measurements that overlap the Pa$\alpha$+24$\mu$m-based estimates from \citet{KENNICUTT07}. They studied selected apertures, while we sample the whole inner disk, so there are methodological differences. But the overall magnitude of both the gas and SFR estimates agree well. 

Finally, note from the first panel that $\tau_{\rm Dep}^{\rm mol}$ estimated using only 70 $\mu$m emission and our adopted scaling agrees very well with that estimated using the four band \citet{GALAMETZ13} fit. That is, the approach that we use at $\theta = 30\arcsec$ agrees well with that which we are forced to use $\theta = 10\arcsec$. The median ratio agrees by construction, but the small scatter gives us confidence in our use of 70 $\mu$m emission and our application of Equation \ref{eq:70totir} \citep[though see][for a more in depth consideration of IR emission as a function of scale]{BOQUIEN16}.

\section{The CO-to-H$_2$ Conversion Factor}

\begin{figure*}
\plotone{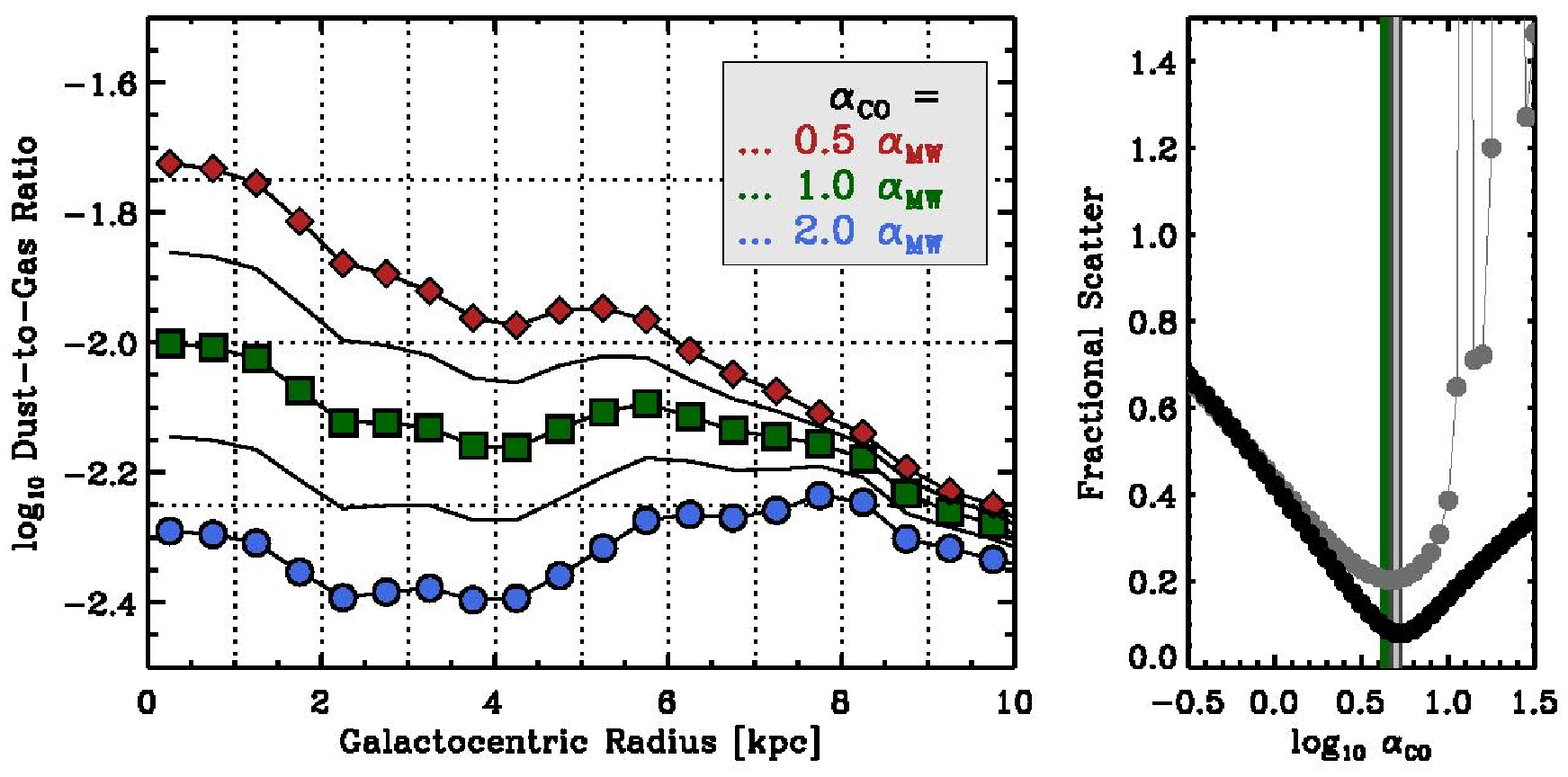}
\caption{Constraints on $\alpha_{\rm CO}$ from comparing CO, {\sc Hi}, and dust mass surface density from fitting the \citet{MENTUCHCOOPER12} {\em Herschel} data using modified versions of the \citet{DRAINE07A} models. ({\em Left:}) The dust to gas ratio as a function of galactocentric radius for different values of $\alpha_{\rm CO}$: from top to bottom the $0.5$ (red), $\sqrt{2}^{-1}$, $1$ (green), $\sqrt{2}$, and $2$ (blue) times the Milky Way $\alpha_{\rm CO} = 4.35$~\acounits . Our adopted Milky Way $\alpha_{\rm CO}$ yields a nearly flat dust-to-gas ratio, consistent with the weak metallicity gradient in the galaxy \cite{CROXALL15}. ({\em Right:}) Solution for $\alpha_{\rm CO}$ (in units of \acounits ) for individual pixels (gray) and radial profile bins (black) assuming a fixed dust-to-gas ratio. We use the fractional minimization technique of \citet{SANDSTROM13} and find the least scatter in the dust to gas ratio for $\alpha_{\rm CO} = 4.5{-}5$~\acounits. For a more detailed analysis using multiple techniques, we refer the reader to B. Groves et al. (in preparation).}
\label{fig:dgraco}
\end{figure*}

We translate CO~(1-0) emission into molecular mass assuming a fixed $\alpha_{\rm CO} = 4.35$~\acounits . This value is supported by multi-line \citep{SCHINNERER10} and cloud virial mass \citep{COLOMBO14A} studies. \citet{SCHINNERER10} provide a thorough summary of the literature on $\alpha_{\rm CO}$ in M51, which has so far yielded results that break down into either an approximately Galactic conversion factor or values $\sim 0.5$ times Galactic. If the lower $\alpha_{\rm CO}$ holds, there would be less molecular gas mass than we infer in the main paper, and a shorter $\tau_{\rm Dep}^{\rm mol}$. This would increase \effsdavg\ by $(\alpha_{\rm CO} / \alpha_{\rm MW})^{-1.5}$, because the conversion factor also affects the density and so $\tau_{\rm ff} \propto 1/ \sqrt{\rho}$.

Figure \ref{fig:dgraco} shows that an approximately Galactic conversion factor is also supported by the dust-based approach of \citet{SANDSTROM13} and \citet{LEROY11}. We compare $\Sigma_{\rm dust}$, the dust mass surface density estimated from {\em Herschel} multi-band data, to the measured CO intensity and the {\sc Hi} column density from VLA imaging. The CO map is the PAWS single dish map, the {\sc Hi} map comes from THINGS \citep{WALTER08}. The dust maps is the result of fitting using the \citet{DRAINE07A,DRAINE07B} models to the {\em Herschel} and {\em Spitzer} photometry, following \citet{ANIANO12} and modified by the correction to dust mass suggested in \citet{PLANCKDRAINE16}.

For this application, we assume that the dust-to-gas ratio is constant over the range $r_{\rm gal} = 1{-}8$~kpc. The approximately constant metallicity of the galaxy supports this assumption \citep[e.g.,][]{CROXALL15}. The figure shows that $\alpha_{\rm CO} \approx \alpha_{\rm MW} = 4.35$~\acounits\ yields an approximately flat dust-to-gas ratio as a function of radius. A lower conversion factor, as suggested by \citet{NAKAI95,WALL16} yields a strong gradient in dust-to-gas ratio as a function of radius. The right panel shows the formal results of minimizing scatter in DGR while varying $\alpha_{\rm CO}$ treating either each ring (black) or each $\theta = 30\arcsec$ line of sight (gray) as independent measurements. Both approaches yield a best fit $\alpha_{\rm CO} \approx 4.5{-}5.0$~\acounits . 

Uncertainties apply to this dust-based approach, including phase- or density-dependent depletion \citep{JENKINS09}, emissivity variations \citep[e.g.,][]{OSSENKOPF94}, and the presence of sufficient dynamic range in the H$_2$/{\sc Hi} ratio to achieve a good fit \citep{SANDSTROM13}. The interplay of these uncertainties with $\alpha_{\rm CO}$ variation are discussed at length in \citet{ISRAEL97}, \citet{LEROY07}, \citet{LEROY11}, \citet{SANDSTROM13}, and \citet{ROMANDUVAL14}, and are beyond the scope of this paper. The key point, for us, is that the best current available dust and gas maps suggest our adopted $\alpha_{\rm CO} \approx \alpha_{\rm MW}$ to represent a reasonable choice.

\end{appendix}

\end{document}